\documentclass{article}

\usepackage{arxiv}

\usepackage[utf8]{inputenc} % allow utf-8 input
\usepackage[T1]{fontenc}    % use 8-bit T1 fonts
\usepackage{hyperref}       % hyperlinks
\usepackage{url}            % simple URL typesetting
\usepackage{booktabs}       % professional-quality tables
\usepackage{amsfonts}       % blackboard math symbols
\usepackage{nicefrac}       % compact symbols for 1/2, etc.
\usepackage{microtype}      % microtypography
\usepackage{lipsum}		% Can be removed after putting your text content
\usepackage{graphicx}
\usepackage{natbib}
\usepackage{doi}

\usepackage{amsthm}
\usepackage{amsmath} 
\usepackage{enumerate}
\usepackage{bbm}
\usepackage{upgreek}
\usepackage{dsfont} % for indicator function
\usepackage{mathtools} % for coloneq
\usepackage{multirow} % for tables
\usepackage{algpseudocode} % for pseudoalgorithm
\usepackage{algorithm} % for pseudoalgorithm

\theoremstyle{remark}

\theoremstyle{definition}

\theoremstyle{remark}

\makeatletter
\newcommand*{\textlabel}[2]{%
  \edef\@currentlabel{#1}% Set target label
  \phantomsection% Correct hyper reference link
  #1\label{#2}% Print and store label
}
\makeatother

\newcommand{\R}{\mathbb{R}}
\newcommand{\N}{\mathbb{N}}

\newcommand{\Thetab}{\boldsymbol{\Theta}}

\title{Multiple change-point detection on the circle via isolation using permutation testing}

\author{\href{https://orcid.org/0009-0006-1372-7032}{\includegraphics[scale=0.06]{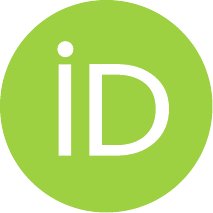}\hspace{1mm}Sophia Loizidou} \\
    Department of Mathematics \\
	University of Luxembourg\\
	\texttt{sophia.loizidou@uni.lu} \\
    \And
    \href{https://orcid.org/0000-0002-7549-1348}{\includegraphics[scale=0.06]{Plots/orcid.pdf}\hspace{1mm}Andreas Anastasiou} \\
    Department of Mathematics and Statistics \\
	University of Cyprus\\
	\texttt{anastasiou.andreas@ucy.ac.cy} \\
	\And
	\href{https://orcid.org/0000-0002-2290-8437}{\includegraphics[scale=0.06]{Plots/orcid.pdf}\hspace{1mm}Christophe Ley} \\
    Department of Mathematics \\
	University of Luxembourg\\
	\texttt{christophe.ley@uni.lu} \\
}

\renewcommand{\shorttitle}{Multiple change-point detection on the circle via isolation using permutation testing}

\begin{document}
\maketitle

\begin{abstract}
In this paper we propose a new method for multiple change-point detection for piecewise-constant circular signals, a setting that, despite its importance in many scientific domains, remains comparatively under-explored. 
The proposed method, Permutation-based Circular Isolate-Detect, denoted PCID, uses an appropriately chosen contrast function and permutation testing to detect change-points in an offline manner, for the data sequence under consideration. 
Prior to detection, PCID isolates the change-points.
The contrast function used is derived under the assumption of von Mises distribution for the noise, but we show that the method is robust and performs well for other distributions as well. 
Simulations are used to showcase the usability of the method in different signal and noise structures, including serially correlated noise.
In order to exhibit the practical relevance of the method in real-world applications, PCID is applied to three real-world datasets, namely flare, acrophase and wave data.
\end{abstract}

% keywords can be removed

\section{Introduction}\label{sec: intro}

Change-point detection, or data segmentation, refers to the task of identifying sudden changes in a dataset where one or more of its characteristics change over time. This problem has drawn considerable attention over the years because of its significance in time series analysis and its broad range of application areas. Among other fields, change-point detection has been employed in neuroscience \citep{seeded_bs}, finance \citep{Cho2012, DAIS}, genomics \citep{genomics1, genomics2}, seismic data \citep{seismic1, seismic2, seismic3} and astronomy \citep{lombaard1991changepoint, LOMBARD1990285, astronomy2}.

Change-point detection on the real line has attracted a lot of interest.
Many papers focus on the problem of detecting changes in the mean of the underlying signal, which is also the scenario considered in this paper, albeit on a different manifold (further explanation can be found in Section~\ref{sec: contrast_function}).
The methods can mainly be  split into two categories, depending on whether the change-points are detected all at once or one by one.
The former category includes optimization based methods with some penalization function to avoid overfitting.
The latter category involves the use of contrast functions, which are applied at each step of the algorithm in order to choose the most probable location of the change-point.
We focus on the second category, to which the method proposed in this paper belongs.
We will only mention some methods here, and refer the interested reader to \cite{review2021, CHO202476} for comprehensive overviews and more detailed explanations.
A method that has received a lot of attention is Binary Segmentation \citep{binary_segmentation}.
It iteratively searches the data for a single change-point, splitting the sequence at each detected location. Because it only searches for one change-point per interval, it can have suboptimal accuracy when multiple changes occur within the same segment, motivating numerous variants of the method.
\cite{WBS} and \cite{WBS2} proposed Wild Binary Segmentation (WBS) and WBS2, respectively, which calculate the value of the contrast function on a large number of randomly drawn intervals that are either drawn once at the beginning (WBS) or every time a detection occurs (WBS2).
\cite{NOT} introduced the Narrowest Over Threshold (NOT) algorithm, which 
chooses the narrowest interval for which the value of the chosen contrast function exceeds a pre-defined threshold, achieving isolation of change-points with high probability, and \cite{seeded_bs} proposed Seeded Binary Segmentation (SeedBS) which uses a deterministic construction of search intervals which can be pre-computed.
Isolate-Detect (ID) \citep{anastasiou2022detecting} achieves isolation of the change-points in a deterministic way before detecting them, while DAIS \citep{DAIS} does so in a data-adaptive way.
As will be explained later in the Introduction, the isolation aspect of the algorithms is important as it offers various advantages.
% The isolation aspect maximises the detection power of the two methods.

A powerful statistical tool that has been used in change-point detection is permutation testing.
It is a versatile technique that is used in many areas of statistics, including hypothesis testing \citep{ernst2004permutation, ge2003resampling}, linear correlation and regression \citep{huh2001random, long2009tetrachoric} and analysis of variance \citep{jung2007new, reiss2010distance}, to name a few.
For a comprehensive historical development of permutation  methods in statistics, see \cite{berry2014chronicle}.
\cite{ANTOCH200137} used permutation testing to get an approximation to the critical values for various tests 
% related to contrast functions 
which are suitable for at most one change-point in the mean or variance.
\cite{huvskova2001permutation} extended such results to multiple changes in the mean and focused on test statistics generated by a kernel function.
\cite{zeileis2013toolbox} embedded permutation testing into various tests for structural changes and
\cite{HORVATH2005351} employed permutations for U-statistics.

An interesting category of data which can be analysed using change-point detection methods are angular data.
% However, change-point detection for angular data is not so well-studied. 
Methods on the real line are not appropriate for such data, as they do not respect their $2\pi$-periodic nature. For example, considering data defined on $[0, 2\pi)$, two consecutive points with observed values at 0 and 6.2 radians are just $2\pi-6.2 \approx 0.08$ radians apart. This can only be recognised by algorithms that understand this periodicity of data points taking values on the unit circle.
Angular data occur naturally in many applications, including animal movement \citep{stephens1969techniques, ghosh_change-point_1999}, wind directions \citep{wind_direction}, health monitoring \citep{Hawkins11062017, LOMBARD2017, Lacomba03052025} and trajectory \citep{Lombard1986}. 
A more comprehensive summary of applications can be found in \cite{ley_modern_2017, LV19}.
It is important to note that we consider circular time-dependent data that involve angles measured during a time period, meaning that the time variable is taken to be linear.
Data for which the time itself is viewed as a circular variable \citep{gill2010circular} are not considered in this work.

Change-point detection on the circle has not been studied as extensively as on the real line.
In what follows we will only mention some methods; for a more comprehensive overview we refer the reader to~\cite{Potgieter2022}.
\cite{grabovsky_change-point_2001} considers a single change in the mean of the von Mises distribution, whose density is given in \eqref{eq: vonMises}, and builds a contrast function based on the log-likelihood ratio. Test statistics are proposed for both known and unknown concentration parameters and they are compared with thresholds obtained from simulations.
Considering the same problem of detecting a change in the mean of the von Mises distribution, \cite{sengupta_bayesian_2008} developed a Bayesian change-point detection method.
\cite{hawkins2015segmentation} employed dynamic programming to calculate the location of multiple change-points using maximum likelihood estimation and the Akaike information criterion as  penalization function.

Some non-parametric methods have also been proposed in the literature.
\cite{Lombard1986} studied non-parametric tests for multiple change-point detection while \cite{CSORGOXB199661} considered rank-based tests for a single change-point.
\cite{grabovsky_change-point_2001} proposed a CUSUM method for a single change-point in the distribution, which is studied for the von Mises as well as other distributions of the observed data.
A more general scenario of random objects with periodic behaviour is considered in \cite{xu_change_2025}.

All aforementioned methods deal with offline change-point detection, meaning that all data are available prior to further analysis.
For completeness, we also mention some methods that focus on analysing data that arrive sequentially in real time, called sequential (or online) detection methods. 
\cite{Gadsden1981} developed sequential probability ratio tests for changes in the mean direction for data that are assumed to follow the von Mises distribution.
Under the same assumption, \cite{LOMBARD2017} proposed sequential CUSUM schemes for detecting changes in the mean direction and concentration, while \cite{lombard2012cusum} proposed sequential CUSUMs to detect a possible change from a uniform distribution on the circle to a non-uniform distribution.
% \textcolor{blue}{\cite{Hawkins11062017} considers distribution-free CUSUMS.}

In this paper we propose an algorithm for offline multiple change-point detection for data on the circle, which we call Permutation-based Circular Isolate-Detect (PCID).
The novelty of the proposed methodology does not come from the specific scenario we are considering, but from the method's broad applicability.
Firstly, PCID is the first proposed algorithm on the circle that attempts to isolate the change-points before detecting them.
Isolating change-points is an important aspect of the algorithm because it offers various advantages.
As highlighted in \cite{anastasiou2022detecting}, change-point detection becomes easier when there is at most one change-point in the interval under consideration.
The estimation accuracy is enhanced by preventing intervals with multiple change-points, allowing the detection of frequent change-points with possibly low-magnitude changes. Moreover, it allows our algorithm to be adapted to higher-order polynomial signals by using the appropriate contrast function
% , as already mentioned, it maximizes \textcolor{blue}{increases, as explained before} the detection power of the contrast function 
(for a discussion on the importance of isolation we refer the reader to Section~\ref{sec: methodology}).
Secondly, our method uses permutation testing as a decision rule.
This means that PCID does not depend on the (asymptotic) distribution of the contrast function used, so our algorithm can be used to detect any type of change as long as an appropriate contrast function is calculated. 
Thirdly, even though our paper focuses on the problem of detecting multiple changes in the mean of the von Mises distribution (see Section~\ref{sec: contrast_function}),
through simulations we show that our algorithm is robust to the underlying distribution of the data under consideration (for more details see Section~\ref{sec: different_noise}).

PCID extends the ID algorithm proposed by \cite{anastasiou2022detecting} to circular data.
As already mentioned, ID first attempts to isolate the change-points, before detecting them using an appropriately chosen contrast function.
In order to achieve isolation, the ID algorithm scans the data sequence in slowly expanding subintervals of $\{1, \ldots, T\}$ 
% of the original sequence, which is defined for time  $t\in\{1, \ldots, T\}$.
% The new intervals 
which are created in a deterministic way, by keeping one of the end-points fixed (at $1$ or $T$) and expanding at each step by an expansion parameter $\lambda_T$.
This way, the $j^\text{th}$ right- and left-expanding intervals are given by $R_j = [1, \min\{j\lambda_T, T\}]$ and $L_j = [\max\{1, T - j\lambda_T + 1\}, T]$, respectively, for $j \in \{1, \ldots, K^{\max} \}$ with $K^{\max} = \lceil T/\lambda_T \rceil $.
The algorithm searches for change-points in each one of the intervals in the ordered set $S_{LR} = \{R_1, L_1, R_2, L_2, \ldots, R_{K^{\max}}, L_{K^{\max}}\}$, using an appropriately chosen contrast function.
As the expansions are performed in a systematic way, which eliminates any randomness in the creation of the intervals, the ID methodology guarantees that every change-point will be isolated in at least one interval, as long as $\lambda_T$ is smaller than the smallest distance between two consecutive change-points.
PCID uses the same strategy as ID, leveraging a contrast function that is suitable for circular data, which we derive in Section~\ref{sec: contrast_function}.
The maximum value of the contrast function for each subinterval is recorded and the algorithm decides if the value obtained is large enough to signify a change-point, using permutation testing. 
% identifies the maximum value that a suitably chosen contrast function obtains (more details on the choice of contrast function can be found in Section~\ref{sec: contrast_function}).
% The algorithm decides if the value obtained is large enough to signify a change-point, using permutation testing.
% PCID uses permutation testing as a decision rule.
% This means that PCID does not depend on the (asymptotic) distribution of the contrast function used, so our algorithm can be used to detect any type of change as long as an appropriate contrast function is calculated. As explained in Section~\ref{sec: contrast_function}, our paper focuses on the problem of detecting multiple changes in the mean of the von Mises distribution.
% However, through simulations we show that our algorithm is robust to the underlying distribution of the data under consideration (for more details see Section~\ref{sec: different_noise}).
% The novelty of our proposal does not come from the specific scenario we are considering, but from the broad applicability of the proposed method.
More details about the algorithm and the permutations performed can be found in Section~\ref{sec: methodology}.

The outline of the paper is as follows. 
In Section~\ref{sec: contrast_function}, the appropriate contrast function is calculated.
The algorithm is explained in Section~\ref{sec: methodology} and a toy example is also given.
A variant of the algorithm in case of long signals is provided in Section~\ref{sec: practicalities}, followed by a discussion on parameter selection.
Simulation results are presented in Section~\ref{sec: simulations}, concerning independent noise structures, more specifically von Mises distribution in Section~\ref{sec: von_mises} and wrapped Cauchy and wrapped Normal distributions in Section~\ref{sec: different_noise}, as well as serially correlated noise in Section~\ref{sec: correlated}.
Three real-data examples are analysed in Section~\ref{sec: real_data}.
The first two examples, which deal with flare and acrophase data, have been analysed before in the literature.
The third dataset concerns wave data and to the best of our knowledge this work is the first to analyse it for change-points.
Section~\ref{sec: conclusions} concludes the paper.
The Appendix contains some further simulation results concerning the choice of the expansion parameter.

\section{Derivation of the contrast function}\label{sec: contrast_function}

In this section, we derive the contrast function which is used for the detection of change-points. 
% in the mean direction of the observed signal.
Our aim is to obtain an expression that is maximised at the same point as the logarithm of the likelihood ratio for all possible single change-point locations within the interval $[s,e)$, for $1\leq s<e\leq T$. 
We work under the model
\begin{equation}\label{eq: model}
    \Theta_t = f_t + \epsilon_t \mod 2\pi, \ t \in \{1,\ldots,T\},
\end{equation}
where $\Theta_t \in [0, 2\pi)$ is the observed data sequence, $f_t \in [0, 2\pi)$ is the unobserved true underlying signal and the independent and identically distributed $\epsilon_t \in [0, 2\pi)$ is the noise.
The scenario we are interested in is piecewise-constant signals, meaning that for $j\in\{1,\ldots,N\}$, $f_t$ satisfies
\begin{align*}
       & f_t = \mu_j \text{ for } t = r_{j-1}+1, \ldots, r_j,\\ 
       & f_{r_j} \neq f_{r_j+1},
\end{align*}
where $1 \leq r_1 < \ldots < r_N < T$ denote the locations of the $N$ change-points and we set $r_0 = 0$.
The hypothesis we are testing is
\begin{align} \label{eq: hypothesis}
    & H_0: f_t = f_{t+1}, \quad \forall t \in \{ 1, \ldots, T-1 \}, \nonumber \\
    & H_A: \exists \ r_1, \ldots, r_N \in \{ 1, \ldots, T-1 \} \text{ such that } f_{r_i} \neq f_{r_i + 1} \ \forall i \in \{1, \ldots, N\}.
\end{align}
Denoting by $\text{vM}(\mu, \kappa)$ the von Mises distribution with location parameter $\mu\in [0, 2\pi)$ and concentration parameter $\kappa\geq0$, 
we assume that $\epsilon_t, t=1,2,\ldots,T$ are independent and identically distributed from the $\text{vM}(0, \kappa)$. The density of the von Mises distribution is given by
\begin{equation}\label{eq: vonMises}
    f(\theta; \mu, \kappa) = \frac{1}{2\pi I_0(\kappa)} \exp \left( \kappa \cos(\theta-\mu) \right)
\end{equation}
where 
\begin{equation}\label{eq: bessel_function}
    I_p(\kappa) = \int_{0}^{2\pi} \cos(p\theta) e^{\kappa \cos(\theta-\mu)} \mathrm{d}\theta
\end{equation}
% $I_0(\kappa) = \int_{0}^{2\pi} \exp \left( \kappa \cos(\theta-\mu) \right) \mathrm{d}\theta$ 
is the modified Bessel function of the first kind and order $p$.
The concentration parameter $\kappa$ controls the variability of the noise and can be understood as the inverse of the variance.
Larger values of $\kappa$ provide more concentrated data around the mean, $\mu$, while $\kappa = 0$ reduces \eqref{eq: vonMises} to the uniform distribution on the circle.
In this paper we assume that $\kappa$ is unknown and constant over time.
Note that, using this assumption on $\epsilon_t$, it is straightforward to show that $\Theta_t \sim \text{vM}(f_t, \kappa)$ for $t=1,\ldots,T$.

The likelihood function of the von Mises distribution $\text{vM}(\mu, \kappa)$ for $\Theta_s, ..., \Theta_e$ can be derived as
\begin{align*}
    L (\Theta_s, \ldots, \Theta_e; \mu, \kappa) 
    & = \left( \frac{1}{2\pi I_0(\kappa)} \right)^{e-s+1} \exp \left( \kappa \sum_{i=s}^e \cos(\Theta_i - \mu) \right)
\end{align*}
and the maximum likelihood estimator (MLE) of the parameter $\mu$ can be calculated to be \citep[(5.3.2)]{mardia_directional_2000}
\begin{equation*}\label{eq: sample_mean}
    \hat{\mu}_{s,e} = \text{atan}2 \left( \sum_{i=s}^e \sin(\Theta_i), \sum_{i=s}^e \cos(\Theta_i) \right),
\end{equation*}
where 
\begin{equation*}
    \text{atan}2 (x, y) = 
    \left\{ \begin{array}{cc}
        \arctan \left( \frac{x}{y} \right) & \text{if } y > 0 \\
        \arctan \left( \frac{x}{y} \right) + \pi & \text{if } y < 0 \\
        \frac{\pi}{2} \text{sign}(x) & \text{if } y = 0, x \neq 0 \\
        0 & \text{if } y = 0, x = 0 \\
    \end{array} \right..
\end{equation*}
The MLE thus, corresponds to the circular sample mean. The log-likelihood ratio statistic of the model under consideration is given by
\begin{align*}
    \textrm{LR} 
    & = 2\log \left( \frac{\sup_{\mu}L(\Theta_s, \ldots, \Theta_b; \mu) \sup_{\mu}L(\Theta_{b+1}, \ldots, \Theta_e; \mu)}{\sup_{\mu}L(\Theta_s, \ldots, \Theta_e; \mu)} \right) \\
    % & = 2 \log \left( \frac{\left( \frac{1}{2\pi I_0(\kappa)} \right)^{b-s+1} \exp \left( \kappa \sum_{i=s}^b \cos(\Theta_i - \hat{\mu}_{s,b}) \right) \left( \frac{1}{2\pi I_0(\kappa)} \right)^{e-b} \exp \left( \kappa \sum_{i=b+1}^e \cos(\Theta_i - \hat{\mu}_{b+1,e}) \right)}{\left( \frac{1}{2\pi I_0(\kappa)} \right)^{e-s+1} \exp \left( \kappa \sum_{i=s}^e \cos(\Theta_i - \hat{\mu}_{s,e}) \right)} \right) \\
    & = 2 \kappa \left( \sum_{i=s}^b \cos(\Theta_i - \hat{\mu}_{s,b}) + \sum_{i=b+1}^e \cos(\Theta_i - \hat{\mu}_{b+1,e}) - \sum_{i=s}^e \cos(\Theta_i - \hat{\mu}_{s,e}) \right),
\end{align*}
where $b$ is the possible change-point location.
From the definition of the mean resultant length $\Bar{R}_{s,e}$ on the interval $[s,e)$
\citep[(2.2.2)]{mardia_directional_2000}, it holds that 
\begin{equation*}
    \Bar{R}_{s,e} \cos(\hat{\mu}_{s,e}) = \frac{1}{e-s+1} \sum_{i=s}^e \cos(\Theta_i), \quad \Bar{R}_{s,e} \sin(\hat{\mu}_{s,e}) = \frac{1}{e-s+1} \sum_{i=s}^e \sin(\Theta_i),
\end{equation*}
so, using trigonometric identities
 we can write
\begin{align*}
    LR = 2\kappa \left\{ (b-s+1) \Bar{R}_{s,b} 
    + (e-b) \Bar{R}_{b+1, e} 
    - (e-s+1) \Bar{R}_{s,e} \right\}.
\end{align*}
Since the concentration parameter $\kappa$ is assumed to be constant, the multiplicative factor $2\kappa$ can be ignored.
Using the fact that
\begin{equation}\label{eq: mean_resultant_length}
    \Bar{R}_{s,e} = \frac{1}{e-s+1} \sqrt{ \left( \sum_{i=s}^e \cos(\Theta_i) \right)^2 + \left( \sum_{i=s}^e \sin(\Theta_i) \right)^2},
\end{equation}
the final form of our contrast function can be written as
\begin{align} \label{eq: contrast_function_final}
    \tilde{C}^b_{s,e}(\Thetab) 
    \coloneq & \left| \sqrt{ \left( \sum_{i=s}^b \cos(\Theta_i) \right)^2 + \left( \sum_{i=s}^b \sin(\Theta_i) \right)^2}
    + \sqrt{ \left( \sum_{i=b+1}^e \cos(\Theta_i) \right)^2 + \left( \sum_{i=b+1}^e \sin(\Theta_i) \right)^2} \right. \nonumber\\
    & \hspace{1cm} \left. - \ \sqrt{ \left( \sum_{i=s}^e \cos(\Theta_i) \right)^2 + \left( \sum_{i=s}^e \sin(\Theta_i) \right)^2} \right|.
\end{align}
As mentioned in the Introduction, \cite{ghosh_change-point_1999} worked under the same assumption of the von Mises distribution and developed tests for detecting a single change in the mean. The contrast function used in the case of known concentration parameter $\kappa$ is the same as \eqref{eq: contrast_function_final}. However, in the case of unknown $\kappa$, a different contrast function was derived.
As already mentioned, in our paper we assume that $\kappa$ is unknown but constant over time, and its exact value is of no interest to us as we use permutation testing.

\section{Methodology} \label{sec: methodology}

In this section we will explain in detail our algorithm PCID and present a toy example at the very end.
We remind the notation $N$ and $r_1, \ldots, r_N$ for the number and locations of the true change-points, and denote by $\hat{N}$, $\hat{r}_1, \ldots, \hat{r}_N$ their estimates.
As already mentioned in the Introduction, the isolation aspect of the algorithm is achieved by leveraging the expansions used in \cite{anastasiou2022detecting}.
The idea of the methodology is to consider gradually larger, overlapping intervals, with one end-point being fixed and the other shifted.
This is done in a deterministic way, by expanding the length of the interval under consideration by an expansion parameter $\lambda_T \in \N\setminus \{ 0 \}$ at each iteration (for more details on the choice of $\lambda_T$ see Section~\ref{sec: parameters}).
The expansions are performed interchangeably, by iteratively expanding once from the left and once from the right at each step.
% In order to increase the speed of the algorithm, these expansions are performed 
% , in such a way that the length of the interval is always a multiple of $\lambda_T$.
% A similar expansion scheme is also explored in \cite{DAIS}, who consider expanding intervals around a point that is chosen in a data-adaptive way.

We consider the interval $[s,e]$, for $1\leq s< e \leq T$.
The left end-point of the right-expanding interval is kept fixed, at point $s$.
Similarly, the right end-point of the left-expanding interval is fixed at point $e$.
Since the expansions occur in such a way that the number of points in the interval is a multiple of $\lambda_T$, the end-points of the intervals under consideration are given by
\begin{align*}\label{eq: interval_endpoints}
     c^r_j = \min \{ s + j \lambda_T - 1, e \}, \quad c^\ell_j = \max \{ e - j \lambda_T + 1, s \},
\end{align*}
for $j = 1,\ldots,K^{\max}_{s,e}$ where $c^r_j$, $c^\ell_j$ are the end-points of the right- and left-expanding intervals, respectively, and $K^{\max}_{s,e} =  \lceil \frac{e-s+1}{\lambda_T} \rceil $.
The intervals are collected in the ordered set
\begin{equation} \label{eq: intervals}
	I_{s,e} = \{ R_{s,e}[1], L_{s,e}[1], R_{s,e}[2], L_{s,e}[2], \ldots, R_{s,e}[K^{\max}_{s,e}], L_{s,e}[K^{\max}_{s,e}]\}
\end{equation}
where $A[j]$ denotes the $j^\text{th}$ element of set $A$ and 
\begin{align*}\label{eq: intervals_left_right}
    R_{s,e} = \{[s,c^r_1], [s,c^r_2], \ldots, [s,c^r_{K^{\max}_{s,e}}]\} , \quad L_{s,e} = \{[c^\ell_1, e], [c^\ell_2, e], \ldots, [c^\ell_{K^{\max}_{s,e}}, e] \}.
\end{align*}
The set $I_{s,e}$ contains the intervals in the order in which they will be checked by PCID for possible change-points.
Such deterministic expansions of the intervals under consideration guarantee isolation of each of the change-points in at least one interval, as long as the value of the expansion parameter, $\lambda_T$, is smaller than the minimum distance between two consecutive change-points.
In such cases, our algorithm has the opportunity to detect the change-point while it is isolated.
% The isolation aspect of the algorithm is important, because it increases the power of the contrast function, $\left| \tilde{C}^b_{s,e}(\Thetab) \right|$, as defined in \eqref{eq: contrast_function_final}.

\begin{figure}[tbp]
\centering
	\includegraphics[height = 5cm, width=7cm]{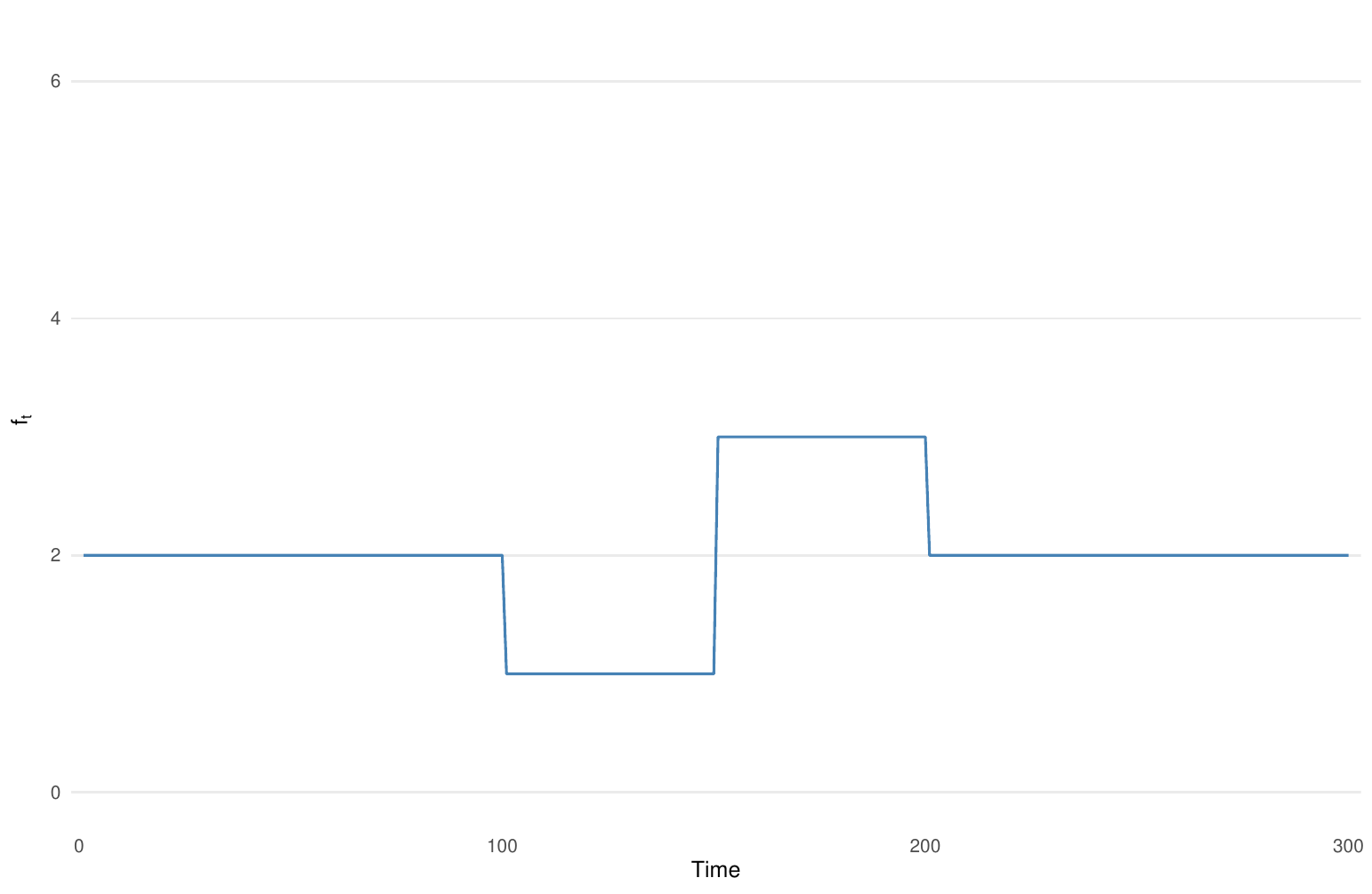}
    \includegraphics[height = 5cm, width=7cm, 
    trim=4cm 0.5cm 4cm 0.5cm, 
    clip]{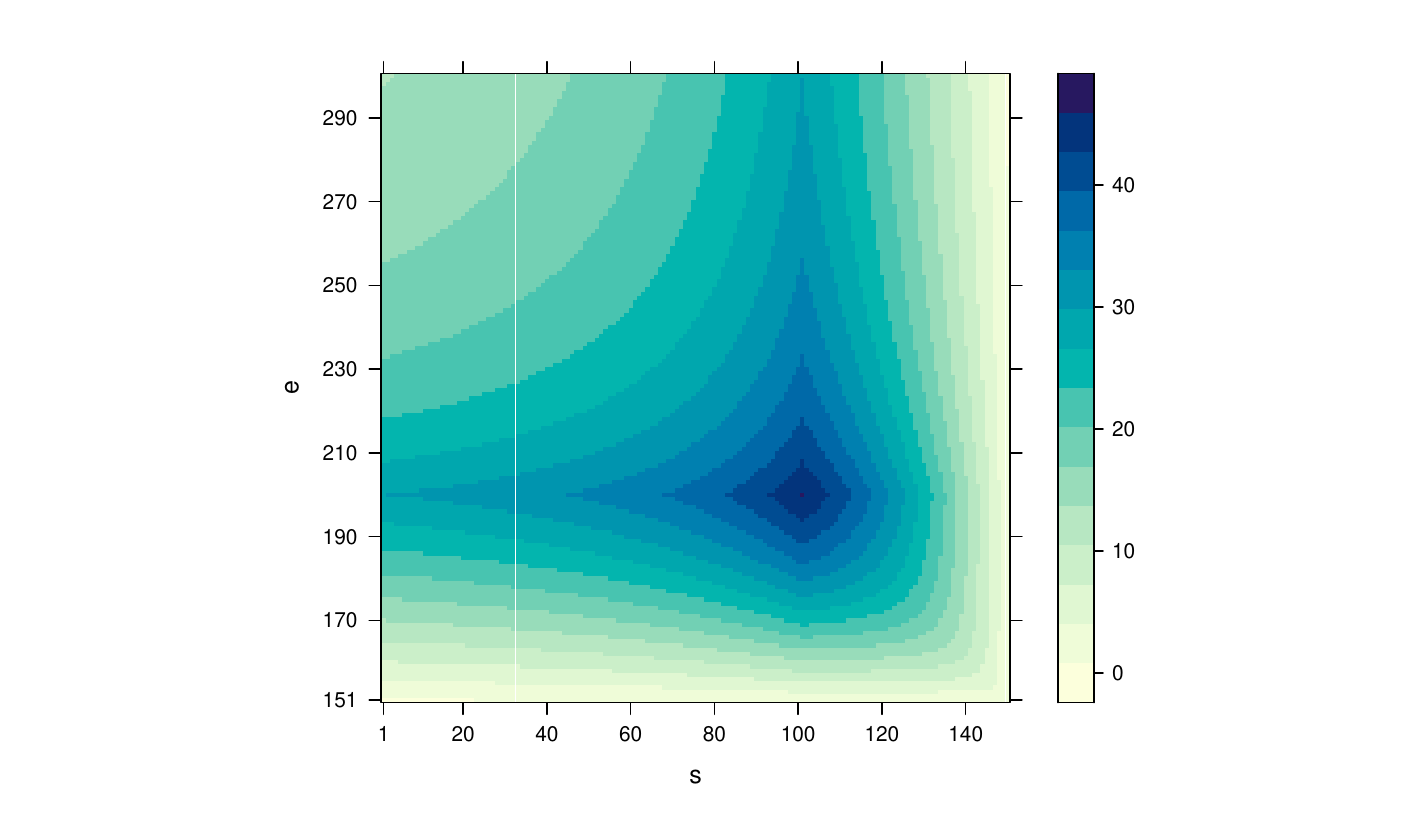} \\
		\hspace*{0.1cm} (a) \hspace{6.25cm} (b) \vspace{0.2cm}
    \caption{(a) Noiseless signal, $f_t$, of length $T=300$ with change-points at locations $100, 150, 200$. (b) Heatmap of the value of the contrast function evaluated on $f_t$ for $b=150$ and all possible combinations of $s=1,\ldots,150$, $e=151,\ldots,300$.}
    \label{fig: heatmap}
\end{figure}

In order to highlight the importance of the isolation of change-points, we consider a noiseless signal, $f_t$, of length $T = 300$.
As shown in Figure~\ref{fig: heatmap}(a), the signal under consideration has 3 change-points at locations $r_1=100, r_2=150, r_3 = 200$ and values between the change-points $2,1,3,2$.
The changes described are, in a sense, opposite, and thus contrast functions have difficulty detecting them.
% Isolating the change-points results in higher detection power and so solves the problem of detecting change-points that cancel each-other out.
In order to exhibit the role of isolation, in Figure~\ref{fig: heatmap}(b) we plot a heatmap of the values that \eqref{eq: contrast_function_final} attains when evaluated on $f_t$ when $b$ is fixed to 150, which is the location of the second change-point.
We consider all possible combinations of $s,e$ taking values in $\{1,\ldots,150\}$ and $\{151,\ldots,300\}$, respectively.
The contrast function achieves the largest value when $s=101, e=200$, which is the largest interval in which $r_2$ is isolated.
This example shows that the value of the contrast function increases when the change-point is isolated, and it is proportional to the length of that interval.
The conclusion is in line with the explanation provided by \cite{DAIS} which states that the value of contrast functions increases as the length of the interval, in which the change-point is isolated, increases.
It is also worth noting that, as explained in \cite{DAIS}, the maximal value is attained when the change-point lays in the midpoint of the interval under consideration, which happens to be the case here.
% This is to be expected as $[101,200]$ is the largest interval in which $r_2$ is isolated.
% This shows that the isolation aspect of the algorithm increases the value of the contrast function, thus increasing its detection power.
% It is also worth noting that, as explained by \cite{DAIS}, contrast functions have the highest power when the change-point lays in the midpoint of the interval under consideration, which happens to be the case here.

In the following paragraphs we will describe how the PCID methodology, presented in Algorithm~\ref{alg: CircID}, detects change-points using permutation testing, as explained in Algorithm~\ref{alg: perm_test}. PCID will check, one by one, the intervals in $I_{s,e}$, in the order in which they appear in \eqref{eq: intervals}.
Without loss of generality we denote the interval under consideration as $[s_1, e_1]\in I_{s,e}$. 
The largest value of the contrast function, denoted by 
\begin{equation}\label{eq: C_obs}
	\tilde{C}_{\text{obs}} = \max_{b \in \{s_1, \ldots, e_1-1\}} \tilde{C}^b_{s_1,e_1}(\Thetab),
\end{equation}
is recorded, and permutation testing is employed to decide if this value is extreme enough to signify the existence of a change-point, as explained in the next paragraph.
If no change-point is detected, the process as just described is repeated for the next interval $[s_2, e_2]\in I_{s,e}$.
If a change-point is detected, its estimated location $\Tilde{b}$ is chosen to be the value where the contrast function is maximized, meaning that $\Tilde{b} = \text{argmax}_{b \in \{s_1, \ldots, e_1-1\}} \tilde{C}^b_{s_1,e_1}(\Thetab) $.
In this case the algorithm restarts using only data included in $[s,\Tilde{b}]$ or $[\Tilde{b}+1, e]$ depending on whether the detection occurred in a left- or right-expanding interval ($[s_1, e_1] \in L_{s,e}$ or $[s_1, e_1] \in R_{s,e}$), respectively. 
%In this case, the data-points that were used to detect $\Tilde{b}$ are discarded, and the algorithm restarts using only data included in $[s,s_1]$ or $[e_1, e]$ if the detection occurred in a left- or right-expanding interval, respectively. 

The permutation testing procedure, as explained in Algorithm~\ref{alg: perm_test}, tests the hypothesis
\begin{align} \label{eq: hypothesis_permutations}
    & H_{0;\, p}: f_t = f_{t+1}, \quad \forall \{ s, \ldots, e-1 \}, \nonumber \\
    & H_{A;\, p}: \exists \ r \in \{ s, \ldots, e-1 \} \text{ such that } f_{r} \neq f_{r } + 1
\end{align}
for an interval $[s,e]$.
Algorithm~\ref{alg: perm_test} resamples the interval under consideration without replacement (so the data points inside the interval are shuffled) at most $B$ times and these permuted intervals are denoted as $P_i(\Thetab; s, e)$, for $i = 1, \ldots, B$.
For each $i$, the maximum value attained by the contrast function on the permuted data, written
\begin{equation*} \label{eq: perm_stat}
    \tilde{C}_{p; i} = \max_{b\in\{s,\ldots,e-1\}} \left\{ \tilde{C}_{s,e}^b(P_i(\Thetab; s, e)) \right\}
\end{equation*}
is recorded.
For the chosen significance level $\alpha_T$, we say that a change-point exists in $[s,e]$ if the number of $\tilde{C}_{p; i}$ that are greater than $\tilde{C}_{\text{obs}}$, as defined in \eqref{eq: C_obs}, does not exceed the cut-off $c_{B,\alpha_T} \coloneq B \times \alpha_T$.
For this reason, in order to reduce the computational complexity of our algorithm permutations are performed up to the point that $\lvert \{ i: \tilde{C}_{p; i} > \tilde{C}_{\text{obs}} \} \rvert = c_{B,\alpha_T}$, where $\lvert A \rvert$ denotes the cardinality of the set $A$.
The algorithm does not perform any permutations of the data sequence $\Theta_s, \ldots, \Theta_e$ if the number of possible permutations $(e-s+1)!$ is less than $B$.

\begin{algorithm}[tbp]
\caption{PCID($\Thetab,s,e,\lambda_T, B, \alpha_T$)}\label{alg: CircID}
\begin{algorithmic}[1]
\If{$e-s<1$}
\State STOP
\Else
\For {$j \in \{1, \ldots, K^{\max}_{s,e}\}$}
\If{no change-point has been detected}
%\State Denote $s_{2j-1} = s$, $e_{2j-1} = c^\ell_j $
%\State Denote $s_{2j} = c^r_j$, $e_{2j} = e $
\State $b_{2j-1} = \text{argmax}_{b\in \{s, \ldots, c^r_j-1\} } \{ \tilde{C}^b_{s, c^r_j}(\Thetab) \}$
\If{Permutation\_test($\Thetab,s,c^r_j, B, \alpha_T, \tilde{C}^{b_{2j-1}}_{s,c^r_j}(\Thetab) $) = TRUE}
\State Add $b_{2j-1}$ to the set of estimated change-points
\State PCID($\Thetab,s,b_{2j-1},\lambda_T, B, \alpha_T$)
\Else{}
\State $b_{2j} = \text{argmax}_{b\in \{c^\ell_j, \ldots, e-1\} } \{ \tilde{C}^b_{c^\ell_j,e}(\Thetab) \}$
\If{Permutation\_test($\Thetab, c^\ell_j, e, B, \alpha_T, \tilde{C}^{b_{2j}}_{c^\ell_j,e}(\Thetab)$) = TRUE}
\State Add $b_{2j}$ to the set of estimated change-points
\State PCID($\Thetab,b_{2j} +1,e,\lambda_T, B, \alpha_T$)
\EndIf
\EndIf
\Else
\State STOP
\EndIf
\EndFor
\EndIf
\end{algorithmic}
\end{algorithm}

\begin{algorithm}[tbp]
\caption{Permutation\_test($\Thetab, s, e, B, \alpha_T, \tilde{C}_{\text{obs}}$)}\label{alg: perm_test}
\begin{algorithmic}[1]
\If{$(e-s+1)! < B$}{}
\State return FALSE
\Else
\State Set $j=0$, $i=1$
\State Set $c_{B,\alpha_T} = B \times \alpha_T$
\While{$i \leq B$ and $j < c_{B,\alpha_T}$}
%\For{$i \in \{1, \ldots, B\}$}
\State $P_i(\Thetab; s, e) = $ Permute($\Theta_s, \ldots, \Theta_e$)
\If{$\tilde{C}_{\text{obs}} > \max_{b\in \{s,\ldots, e-1\}} \tilde{C}^b_{s,e}(P_i(\Thetab; s, e))$} {}
\State $j = j+1$
\EndIf
\State i = i + 1
%\EndFor
\EndWhile
%\If{$ j > B \times \alpha_T$} {}
\If{$ j < c_{B,\alpha_T} $} {}
\State $\Theta_s,\ldots, \Theta_e$ contains a change-point
\State return TRUE
\Else
\State return FALSE
\EndIf
\EndIf
\end{algorithmic}
\end{algorithm}

\begin{figure}[tbp]
    \centering
    \includegraphics[width=1\linewidth]{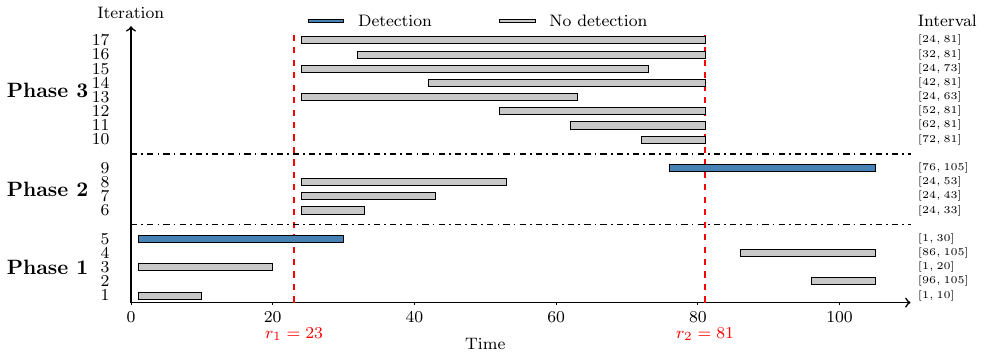}
    \caption{Sequence of length $T=105$ with two change-points at $r_1 = 23$ and $r_2 = 81$.
    The intervals are checked in the order they appear, from the bottom to the top.
    Blue intervals indicate that detection of a change-point has occurred.}
    \label{fig: toy_example}
\end{figure}

We now provide a toy example of how the algorithm works in practice, which is also illustrated in Figure~\ref{fig: toy_example}.
We consider a sequence of length $T=105$ with two change-points, at positions $r_1 = 23$ and $r_2 = 81$, and values of $f_t$ between these change-points of $0,2,0$.
We set $\epsilon_t \sim \text{vM}(0, 2)$
% The random variable $\epsilon_t$ is set to have concentration parameter $\kappa=2$ 
and the values of the parameters are chosen as $\lambda_T = 10$, $\alpha_T = 0.001$ and $B=1000$ for the sake of illustration.
As a first step, the algorithm checks whether the interval $[1,10]$ contains any change-points, by performing permutation testing.
As there are no change-points, in Phase 1 PCID considers the intervals $[96,105], [1,20], [86,105]$ and $[1,30]$, in this order.
A change-point is detected at location $\hat{r}_1 = 23$, so the algorithm restarts on $[24,105]$.
As $[96,105]$ and $[86,105]$ have already been checked, in Phase 2 PCID looks for change-points in $[24,33], [24,43], [24,53]$ and $[76,105]$.
The last interval indicates the existence of a change-point at location $\hat{r}_2 = 81$, and the process is restarted on $[24, 81]$.
Since in Phase 2 the intervals $[24,33], [24,43], [24,53]$ have already been considered,  in Phase 3 PCID only searches for change-points in $[72,81], [62,81], [52,81], [24,63], [42,81], [24,73], [32,81], [24,81]$, specifically with that order. No change-points are detected and thus, the algorithm concludes.

\section{Practicalities} \label{sec: practicalities}

\subsection{Considerations for long signals} \label{sec: variants}

In this subsection we introduce a modification of the algorithm as described in Section~\ref{sec: methodology}, in order to tackle the case of rather long signals. 
Since PCID is based on permutation testing, the computational complexity can increase dramatically when the length of the subinterval under consideration is, even moderately, large.
In order to avoid this issue, when the length of the data sequence exceeds $w$, 
% (see Section~\ref{sec: parameters} for the choice of $w$), 
the sequence is split into disjoint intervals of length at most $w$. More specifically, for a sequence of length $T$, the PCID algorithm is applied on $k_w = \lceil T/w \rceil$ disjoint intervals of the form
\begin{equation*}
% \label{eq: intervals_long_signals}
    J_{i} = [a_{i}, b_{i}] = [w \cdot (i-1) + 1 , \min \{ w \cdot i, T\}], \quad \text{for } i \in \{1,\ldots,k_w\}.
\end{equation*}
After detecting all change-points and in order to ensure that no change-points are left undetected close to the right end-point of interval $J_1$, left and right end-points of $J_i$ for $i\in\{2,\ldots,k_w-1\}$ and the left end-point of $J_{k_w}$,
% end-points of the intervals $J_{i}$, 
we also we carry permutation testing on the intervals given by
\begin{equation*}
    [\max \{ \hat{r}_{i;\hat{N}_i} \mathds{1}_{\{\hat{N}_i>0\}} + 1, b_{i} - \lfloor w/2 \rfloor \}, \min\{ \hat{r}_{i+1;1} \mathds{1}_{\{\hat{N}_{i+1}>0\}}, b_{i} + \lfloor w/2 \rfloor \}], \quad \text{for } i \in \{1,\ldots,k_w - 1\},
\end{equation*}
where we denote by $\hat{N}_i$ and $a_{i} \leq \hat{r}_{i; 1}<\ldots< \hat{r}_{i; \hat{N}_i} < b_{i}$ the number and locations of the detected change-points in $J_{i}$.
In other words, on the one hand, if no change-points have been detected in $J_{i}$ or $J_{i+1}$, we apply Algorithm~\ref{alg: perm_test} on the interval $[b_{i} - \lfloor w/2 \rfloor, b_{i} + \lfloor w/2 \rfloor]$ which has length $2 \lfloor w/2 \rfloor + 1$.
% with the boundaries, $b_{i}$ and $a_{i+1}$, separating the disjoint intervals being in the middle.
% Note that it holds that $a_{i+1} = b_{i} + 1$.
On the other hand, if any change-points have been detected, we choose the largest possible interval of length at most $w + 1$ in which no detected change-points are included.

% The parameter $w$ is introduced in the variant of the algorithm described in Section~\ref{sec: variants} in order to reduce the computational complexity of PCID in the case of long signals.
Considering the choice of $w$, choosing a small value for $w$ defeats the purpose of the variant, as even small data sequences will be split into smaller ones. 
However, too large values might not lead to a considerable reduction of computational complexity, as large intervals are still permuted. 
Simulations have shown that a value that is manageable computationally, and provides a considerable computational edge, is $w=500$, which we use in practice.
% ; see Table \ref{table: sims_window} for a comparison between the window- and the non-window-based PCID versions.
% Simulations have shown that the computational complexity of our algorithm for sequences up to length $T=500$ is less than 30s on average, even for difficult signals.

In order to exhibit the reduction of the computational cost using this variant of our algorithm, we present some simulation results in Table~\ref{table: sims_window}.
The signals used for the simulations are the following:
\begin{itemize}
    \item[\textlabel{(S1)}{signal: nocpt_very_long}]  sequence of length $T=1000$ with no change-points.

    \item[\textlabel{(S2)}{signal: cpts_window}] sequence of length $T=1000$ with 4 change-points at locations $200, 400, 600, 800$ and values between change-points $0, 3, 0, 3, 0, 3$.
\end{itemize}
We consider model \eqref{eq: model} when $\epsilon_t \sim \text{vM}(0, \kappa)$, with $\kappa =2$ for Signal~\ref{signal: nocpt_very_long} and $\kappa =4$ Signal~\ref{signal: cpts_window}.
We report the frequency of $\hat{N} - N$ as a measure of accuracy of the number of detected change-points; we remind the reader that $N$ and $\hat{N}$ are the numbers of true and estimated change-points, respectively. 
We also report the Adjusted Rand Index (ARI), ideally close to 1, and the scaled Hausdorff distance $d_H$, ideally close to 0, as measures of accuracy of the location of the detected change-points.
For more details on these measures, we refer the reader to Section~\ref{sec: framework}.
We denote the algorithm using the variant as just described by PCID$_\text{W}$, where `W' stands for `Window'.
% and the one not using the variant as PCID$_\text{NW}$, which stands for `No Window'.
We set $B=1000$ and $\alpha_T = 0.001$ for both PCID and PCID$_\text{W}$, while the biggest window to be considered in PCID$_\text{W}$ is set to $w = 500$.
The results in Table~\ref{table: sims_window} indicate that the proposed variant of the algorithm decreases the computational complexity.
This is especially evident when no change-points are present, since in such cases the intervals on which permutation tests are performed become very large.
Given the chosen value of $w = 500$ and the length of the signals, the largest interval for PCID$_\text{W}$ is of length 500 while for PCID it is 1000.
As will be explained in Section~\ref{sec: parameters}, the choice of parameters $B=1000$ and $\alpha_T = 0.001$ corresponds to around $5\%$ significance level of the algorithm for data sequences of length $T=500$.
Both Signals~\ref{signal: nocpt_very_long} and \ref{signal: cpts_window} have length $T = 1000$, thus, the overestimation of around $10\%$ of change-points that is observed in the table can be understood.
This can be avoided by increasing the value of $B$ and decreasing the value of $\alpha_T$ to allow for a lower significance level, for example taking $B = 10,000$ and $\alpha_T = 0.0001$.
More details on the choice of the parameters are given in Section~\ref{sec: parameters}.
% For the rest of the paper, we use the variant of the algorithm as described in this section.

\begin{table}[tbp]
\caption{Distribution of $\hat{N} - N$ over 100 simulated data sequences of the Signals \ref{signal: nocpt_very_long} and \ref{signal: cpts_window} with $\epsilon_t \sim \text{vM}(0, \kappa)$. 
    The average ARI, $d_H$ and computational times are also given.} \label{table: sims_window}
\centering
        \begin{tabular}{|c|c|l|c|c|c|c|c|c|c|c|c|c|}
            \hline
            % & & & \multicolumn{7}{c|}{} & & & \\ 
            & & & \multicolumn{7}{c|}{$\hat{N} - N$} & & & \\
            Signal & $\kappa$ & Algorithm & $\leq -3$ & -2 & -1 & 0 & 1 & 2 & $\geq 3$ & $d_H$ & ARI & Time (s) \\
            \hline
            \multirow{2}{*}{\ref{signal: nocpt_very_long}} & \multirow{2}{*}{2} & PCID$_\text{W}$ & - & - & - & 89 & 4 & 7 & 0 & - & - & 18.0 \\
            \cline{3-13}
            & & PCID & - & - & - & 86 & 1 & 13 & 0 & - & - & 37.0 \\
            \hline
            \multirow{2}{*}{\ref{signal: cpts_window}} & \multirow{2}{*}{4} & PCID$_\text{W}$ & 0 & 0 & 0 & 91 & 7 & 1 & 1 & 0.007 & 0.995 & 17.6 \\
            \cline{3-13}
            & & PCID & 0 & 0 & 0 & 87 & 8 & 4 & 1 & 0.014 & 0.994 & 21.3 \\
            \hline
\end{tabular}
\end{table}

\subsection{Parameter selection} \label{sec: parameters}

In this subsection, we provide information on the choice of 
% the window length, $w$, 
the expansion parameter, $\lambda_T$, the number of permutations performed, $B$, and the significance level for the permutation tests defined in \eqref{eq: hypothesis_permutations}, $\alpha_T$.

The choice of the value of $\lambda_T$ is not straightforward.
In cases of signals with no change-points, since no detections occur and the algorithm does not restart on smaller intervals, large values of $\lambda_T$ imply that less intervals are searched for change-points, while smaller values will force the algorithm to check more intervals. 
For example, $\lambda_T$ means checking the double amount of intervals compared to $2\lambda_T$, with half of them being all those checked for $2\lambda_T$, due to the nature of the expansions as described in Section~\ref{sec: methodology}.
In cases of signals with change-points, larger values of $\lambda_T$ might be problematic, as one of the biggest advantages of our algorithm, isolation of the change-points, might not be possible, due to $\lambda_T$ being possibly greater than the minimum distance between consecutive change-points. 
%This will be the case when $\lambda_T$ is bigger than the distance between consecutive change-points. 
In Appendix~\ref{sec: lambda_sims} we present some simulation results for the average computational time of PCID for three data sequences; we vary the number of change-points and the distance between them. 
The results are presented for $\lambda_T \in \{2,3,4,5,10,20,30,40,50\}$ and support our explanation for the role of $\lambda_T$ in the computational complexity of the algorithm.
In practice, both for the synthetic as well as the real-world data results, we use $\lambda_T = 5$, which is not too computationally expensive even for long signals, thanks to the variant described in Section~\ref{sec: variants}.

% \todo[inline]{For $\alpha$ say that we use the table to find the closest value but if the value of alpha forces B to be 10.000 then we use $\alpha=0.001$, $B=1000$ to have faster calculation. 
% The user can choose to override this and use $B=10000$.
% We still need $B\times\alpha \in \N$.
% In the case that the desired Type 1 error is very low, we go with the lowest value from the table}

% \todo[inline]{Explain that for our choice of $\lambda_T$, the first interval is not checked as length! less than B}

The number of permutations performed, $B$, plays an important role in the computational time of the method.
Larger values of $B$ improve the accuracy and stability of the estimated p-value of the hypothesis test given in \eqref{eq: hypothesis_permutations} in each repetition of the permutation test and also allow for smaller values of $\alpha_T$, making the algorithm more conservative in detecting change-points. 
However, this comes at the cost of higher computational time.
The value of $B$ is set such that the product $B\times\alpha_T$ is an integer.
The reason for this is to have an integer-valued cut-off $c_{B,\alpha_T} = B\times\alpha_T$.
An alternative would be to consider $\lceil B\times\alpha_T \rceil$, but this would imply that different values of $\alpha_T$ provide the same significance level for test \eqref{eq: hypothesis_permutations}, for example, $B = 1000$ and $\alpha_T = 0.001$ or $\alpha_T = 10^{-10}$.
For the reasons explained, the value of $B$ is set to $B=10^d$, where $d$ is the number of decimals of $\alpha_T$.

The role of the parameter $\alpha_T$ is to control the significance level of the permutation tests performed within the PCID algorithm, under the null hypothesis of no change-points.
However, our real goal is to control the Type I error of the whole algorithm, as described in \eqref{eq: hypothesis} and denoted by $\gamma$, and not the one of each permutation testing procedure separately.
Due to the nature of the algorithm, the exact number of permutation tests performed cannot be calculated in advance. 
For this reason, we ran a large-scale simulation study in order to calculate the Monte Carlo estimate of $\gamma$ for our method, which we denote by $\hat{\gamma}$.
In Table~\ref{tab: alpha}, we present the values of $\hat{\gamma}$ over $1000$ simulations for different values of $\alpha_T$ and signals of length $T=50,100,150,\ldots,500$ without any change-points, since we wish to obtain the rejection frequency under the null hypothesis \eqref{eq: hypothesis}. 
Due to the variant of Section~\ref{sec: variants} and the choice of $w=500$, we only calculate the values up to $T=500$.
In the case that $T>w$ and the data sequence is split into $k_w = \lceil T/w\rceil$ disjoint intervals, we account for multiple hypothesis testing by employing the \v{S}id{\'a}k correction \citep{vsidak1967rectangular}, meaning that we set the Type I error for each of the disjoint intervals to be
\begin{equation}\label{eq: gamma_i}
    \gamma_i \coloneq 1-(1-\gamma)^{1/k_w}
\end{equation}
for $i \in\{1,\ldots,k_w\}$. 
For a given $\gamma$, the value of $\alpha_T$ is chosen as the value from Table~\ref{tab: alpha} that corresponds to the closest $\hat{\gamma}$ according to the length of the sequence rounded to the nearest multiple of 50.
In the case that the PCID$_\text{W}$ variant is employed, a different value of $\alpha_T$ is chosen for each interval, based on the values of $\gamma_i$ and the length of each interval.
% For the chosen Type I error ($\alpha'$ or $\alpha'_i$) our algorithm chooses the value of $\alpha_T$ from Table~\ref{tab: alpha} that corresponds to the closest simulated Type I error according to the length of the sequence rounded to the nearest multiple of 50.
In order to reduce the number of permutations performed, if the value chosen using Table~\ref{tab: alpha} is $\alpha_T < 0.001$, PCID will set $\alpha_T = 0.001$ and $B=1000$.
The user can choose to override this setting in case the calculation needs to be done with a very small significance level, keeping in mind that the calculation will be much slower.

\begin{table}[tbp]
    \centering
    \caption{Monte Carlo estimates of the Type I error of PCID, $\hat{\gamma}$, calculated over $1000$ simulations for different values of $\alpha_T$ and data sequences of length $T=50,100,150,\ldots,500$ with no change-points.
    We set $\lambda_T = 5$, $B=10,000$ and indicate in bold the values of $\hat{\gamma}$ closest to 0.01 and 0.05.}
    \label{tab: alpha}
    \begin{tabular}{c|c|c||c|c|c}
        $T$ & $\alpha_T$ & $\hat{\gamma}$ & $T$ & $\alpha_T$ & $\hat{\gamma}$ \\
        \hline
        \multirow{12}{*}{50} & 0.01 & 0.083 & \multirow{12}{*}{100} & 0.01 & 0.149  \\
         & 0.009 & 0.078 &  & 0.005 & 0.083 \\
         & 0.008 & 0.066 &  & 0.004 & 0.069 \\
         & 0.007 & 0.058 &  & \textbf{0.003} & \textbf{0.051} \\
         & \textbf{0.006} & \textbf{0.046} &  & 0.002 & 0.037 \\
         & 0.005 & 0.041 &  & \textbf{0.001} & \textbf{0.011} \\
         & 0.004 & 0.035 &  & 0.0005 & 0.005 \\
         & 0.003 & 0.029 &  & 0.0001 & 0.001 \\
         & \textbf{0.002} & \textbf{0.008} &  & - & - \\
         & 0.001 & 0.006 &  & - & - \\
         & 0.0005 & 0.002 &  & - & - \\
         & 0.0001 & 0.000 &  & - & - \\
        \hline
        \multirow{7}{*}{150} & 0.005 & 0.097 & \multirow{7}{*}{200} & 0.005 & 0.131 \\
         & \textbf{0.003} & \textbf{0.055} &  & \textbf{0.002} &  \textbf{0.057} \\
        & 0.002  & 0.032 &  & 0.001 & 0.037 \\
        & 0.001  & 0.017 &  & 0.0005 & 0.017 \\
        & \textbf{0.0005} & \textbf{0.010} &  & \textbf{0.0003} & \textbf{0.013} \\
         & 0.0001  & 0.003 &  & 0.0002 & 0.004 \\
         & - & - &  & 0.0001 & 0.003 \\
        \hline
         \multirow{7}{*}{250} & \textbf{0.002} & \textbf{0.056} & \multirow{7}{*}{300} & 0.002 & 0.070 \\
         & 0.001 & 0.034 &  & \textbf{0.001} & \textbf{0.041} \\
         & 0.0005 & 0.019 &  & 0.0005 & 0.021 \\
         & 0.0004 & 0.014 &  & 0.0004 & 0.017 \\
         & 0.0003 & 0.012 &  & 0.0003 & 0.013 \\
         & \textbf{0.0002} & \textbf{0.010} &  & \textbf{0.0002} & \textbf{0.009} \\
         & 0.0001 & 0.002 &  & 0.0001 & 0.003 \\
        \hline
        \multirow{7}{*}{350} & 0.002 & 0.068 & \multirow{7}{*}{400} & 0.002 & 0.076 \\
         & \textbf{0.001} & \textbf{0.044} &  & \textbf{0.001} & \textbf{0.045} \\
         & 0.0005 & 0.019 &  & 0.0005 & 0.025 \\
         & 0.0004 & 0.018 &  & 0.0004 & 0.021 \\
         & 0.0003 & 0.013 &  & \textbf{0.0003} & \textbf{0.013} \\
         & \textbf{0.0002} & \textbf{0.008} &  & 0.0002 & 0.006 \\
         & 0.0001 & 0.007 &  & 0.0001 & 0.003 \\
        \hline
        \multirow{7}{*}{450} & 0.002 & 0.081 & \multirow{7}{*}{500} & 0.002 & 0.096 \\
         & \textbf{0.001} & \textbf{0.048} &  & \textbf{0.001} & \textbf{0.057} \\
         & 0.0005 & 0.020 &  & 0.0005 & 0.031 \\
         & 0.0004 & 0.025 &  & 0.0004 & 0.028  \\
         & 0.0003 & 0.013 &  & 0.0003 & 0.020 \\
         & \textbf{0.0002} & \textbf{0.009} &  & \textbf{0.0002} & \textbf{0.009} \\
         & 0.0001 & 0.005 &  & 0.0001 & 0.002 \\
    \end{tabular}
\end{table}

\section{Simulations}\label{sec: simulations}

\subsection{Framework} \label{sec: framework}

In this section we investigate the performance of PCID for various signal and noise structures. 
% We start by depicting our framework in Section~\ref{sec: framework}. 
It is noteworthy that, to the best of our knowledge, there are no implemented algorithms for multiple change-point detection on the circle, thus we do not compare the performance of PCID with competitors and hence do not describe any competitors.
In the case of large signals, and in order to reduce the computational complexity of our algorithm, we employ the PCID$_\text{W}$ variant with $w=500$, as described in Section~\ref{sec: variants}.
The parameters of PCID are set to the values proposed in Section~\ref{sec: parameters}.
More specifically, $\lambda_T = 5$ and $\alpha_T$ is chosen from Table~\ref{tab: alpha}.
The smallest value of $\alpha_T$ we use is 0.001 and we set $B=1000$. 
% Simulations using smaller values of $\alpha_T$ and $B=10,000$ are provided in Appendix~\ref{sec: further_sims}.
We report the results for targeted Type I error of the test described in \eqref{eq: hypothesis} taking values $\gamma = 0.01$ and $0.05$.
% As the length of all signals presented is less than $w=500$, there is no need to employ the variant of PCID as described in Section~\ref{sec: variants}.
The code for the simulations is available at \url{https://github.com/Sophia-Loizidou/PCID}.

In our results we report the frequency of $\hat{N} - N$ as a measure of accuracy of the number of detected change-points.
As a measure of accuracy of their location, we report the Adjusted Rand Index (ARI) of the estimated segmentation against the true one \citep{hubert1985comparing}. 
We also report the scaled Hausdorff distance, defined as
\begin{equation*}
    d_H = n_s^{-1} \max {\Big\{\max_j \min_k \left| r_j - \hat{r}_k \right|, \max_k \min_j \left| r_j - \hat{r}_k \right| \Big\}}
\end{equation*}
where $n_s$ is the length of the largest segment between successive change-points, $r_j$ are the true change-points and $\hat{r}_j$ their estimates. 
A good performing method should achieve ARI close to 1 and $d_H$ close to 0.
Finally, we also report the average computational time.

The piecewise-constant signals $f_T$ used in the simulations are the following:
\begin{itemize}
    \item[\textlabel{(S3)}{signal: nocpt}]  sequence of length $T=200$ with no change-points.  

    \item[\textlabel{(S4)}{signal: simple}]  sequence of length $T=100$ with 1 change-point at location $50$ and values between change-points $0$ and $\pi$.
    
    \item[\textlabel{(S5)}{signal: simple2}]  sequence of length $T=200$ with 2 change-points at locations $50$ and $100$ with values between change-points equal to $0$, $\pi$ and $1$. 

    \item[\textlabel{(S6)}{signal: simple3}]  sequence of length $T=210$ with 6 change-points at locations $30, 60, 90, 120, 150$, and $180$ with values between change-points equal to $0, 1, 2, 3, 4, 5$, and $6$. 

    \item[\textlabel{(S7)}{signal: complex1}]  sequence of length $T=150$ with 3 change-points at locations $60,100$, and $130$ with values between change-points equal to $1.5, 3.3, 5.2,$ and $1.5$. 

    \item[\textlabel{(S8)}{signal: long}]  sequence of length $T=600$ with 3 change-points at locations $150,300$, and $500$ with values between change-points equal to $1, 4, 2,$ and $5$. 
\end{itemize}
These piecewise-constant signals will be combined with various noises. 
In Section~\ref{sec: independent_noise} we consider the situations where the independent noise $\epsilon_t$ follows the von Mises distribution, $\epsilon_t \sim \text{vM}(0, \kappa)$ (Section~\ref{sec: von_mises}), under which the contrast function is constructed, as well as other circular distributions, namely the wrapped Cauchy, $\epsilon_t \sim \text{wC}(0, \rho)$ and wrapped Normal distributions, $\epsilon_t \sim \text{wN}(0, \beta)$ (Section~\ref{sec: different_noise}).
Section~\ref{sec: correlated} contains settings with dependent noise $\epsilon_t$.
Different values of the concentration parameters $\kappa$, $\rho$ and $\beta$ are considered for Signals \ref{signal: simple} - \ref{signal: long}.
Signal~\ref{signal: nocpt} contains no change-points, thus we only consider one value for the concentration parameter.

\begin{figure}[htbp]
	\includegraphics[height = 5cm, width=6.5cm]{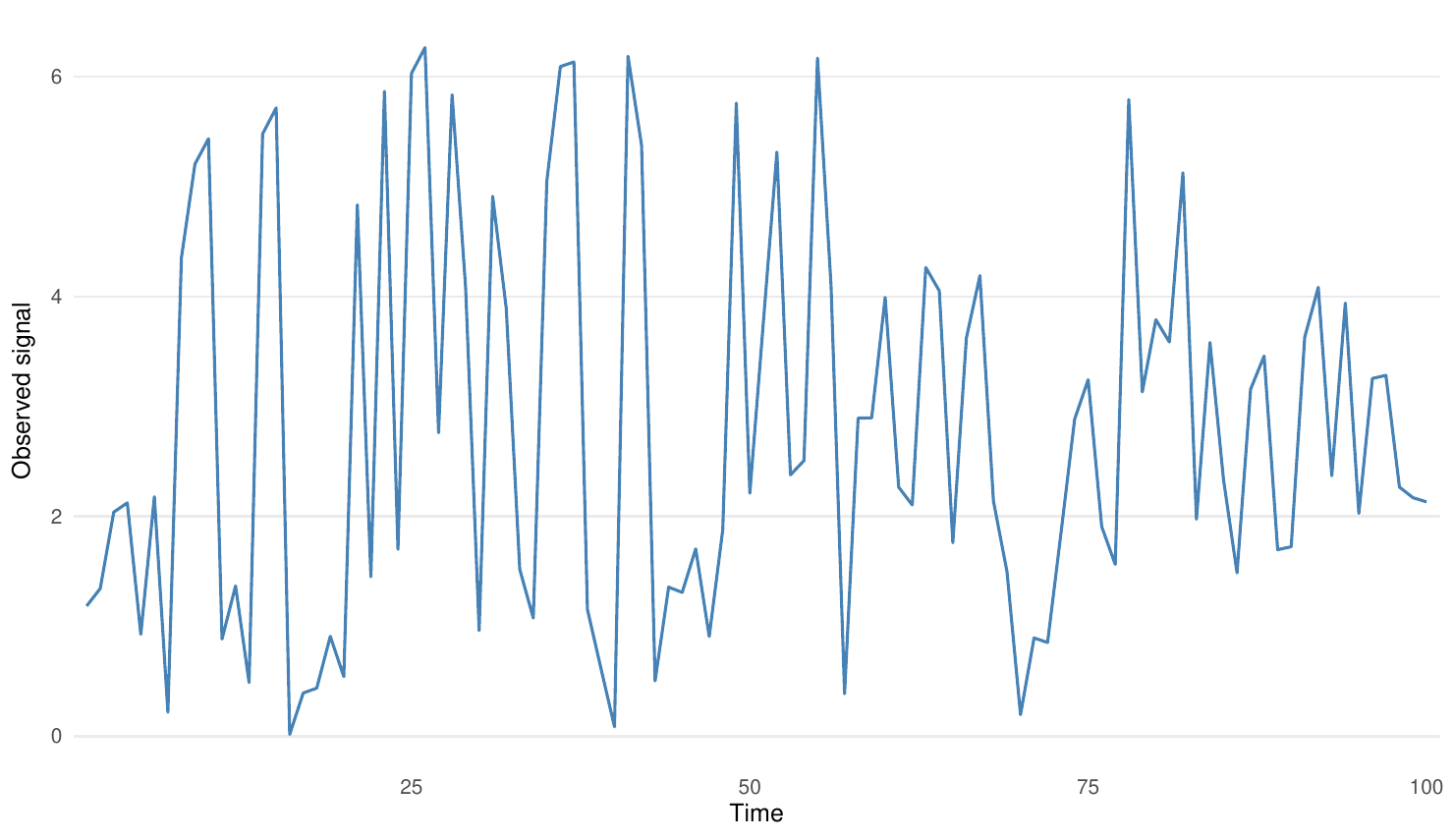} \hspace{0.7cm}
    \includegraphics[height = 5cm, width=7cm]{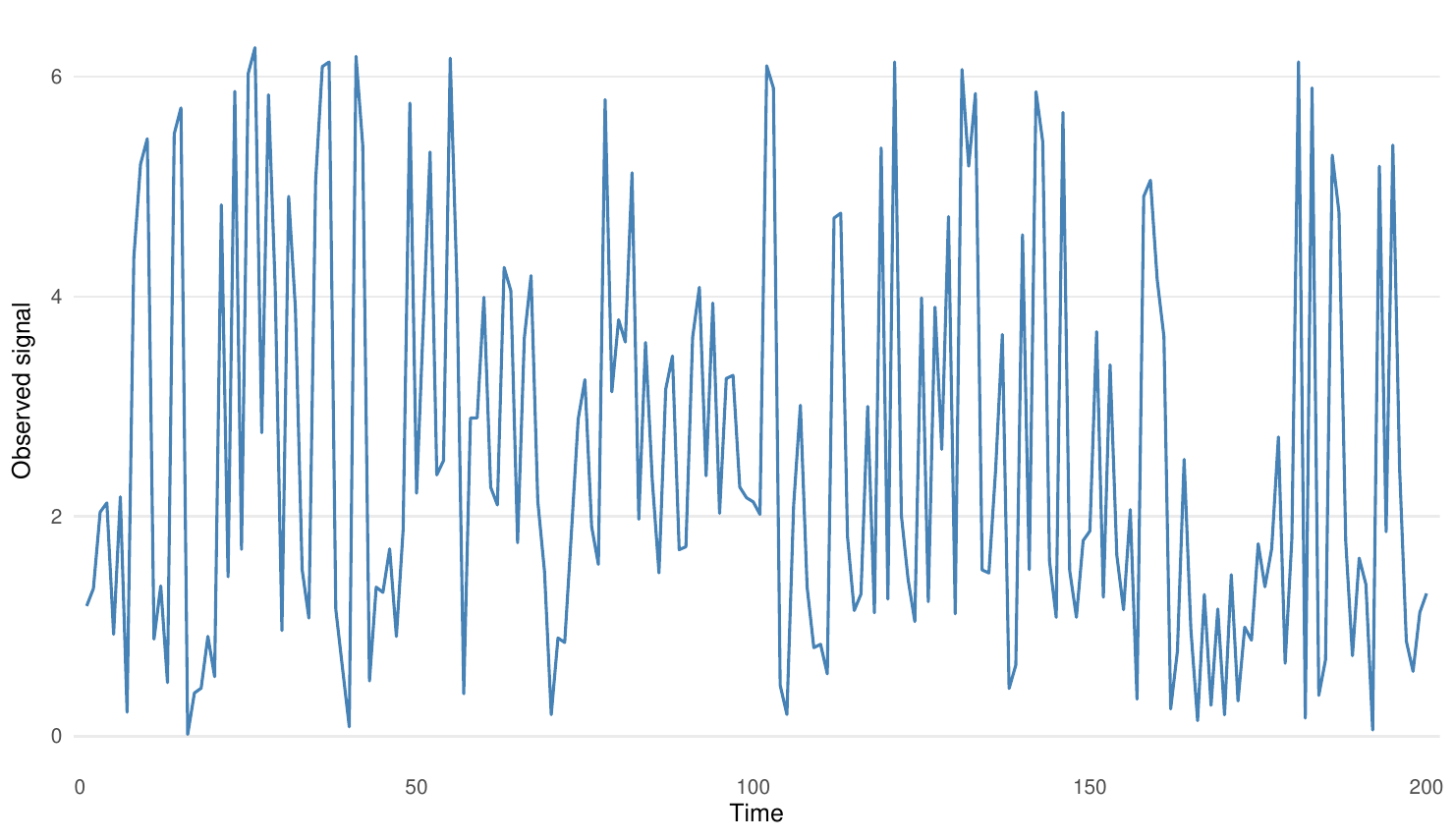} \\
		\hspace*{3.1cm} (a) \hspace{7.05cm} (b) \vspace{0.2cm}
    \caption{Plots of data sequences that follow model \eqref{eq: model} with $\epsilon_t \sim \text{vM}(0, \kappa)$ for $\kappa = 1$ with underlying signals (a) \ref{signal: simple} and (b) \ref{signal: simple2}.}
    \label{fig: signal_plot_simple_simple2}
\end{figure}

\begin{figure}[htbp]
	\includegraphics[height = 5cm, width=6.5cm]{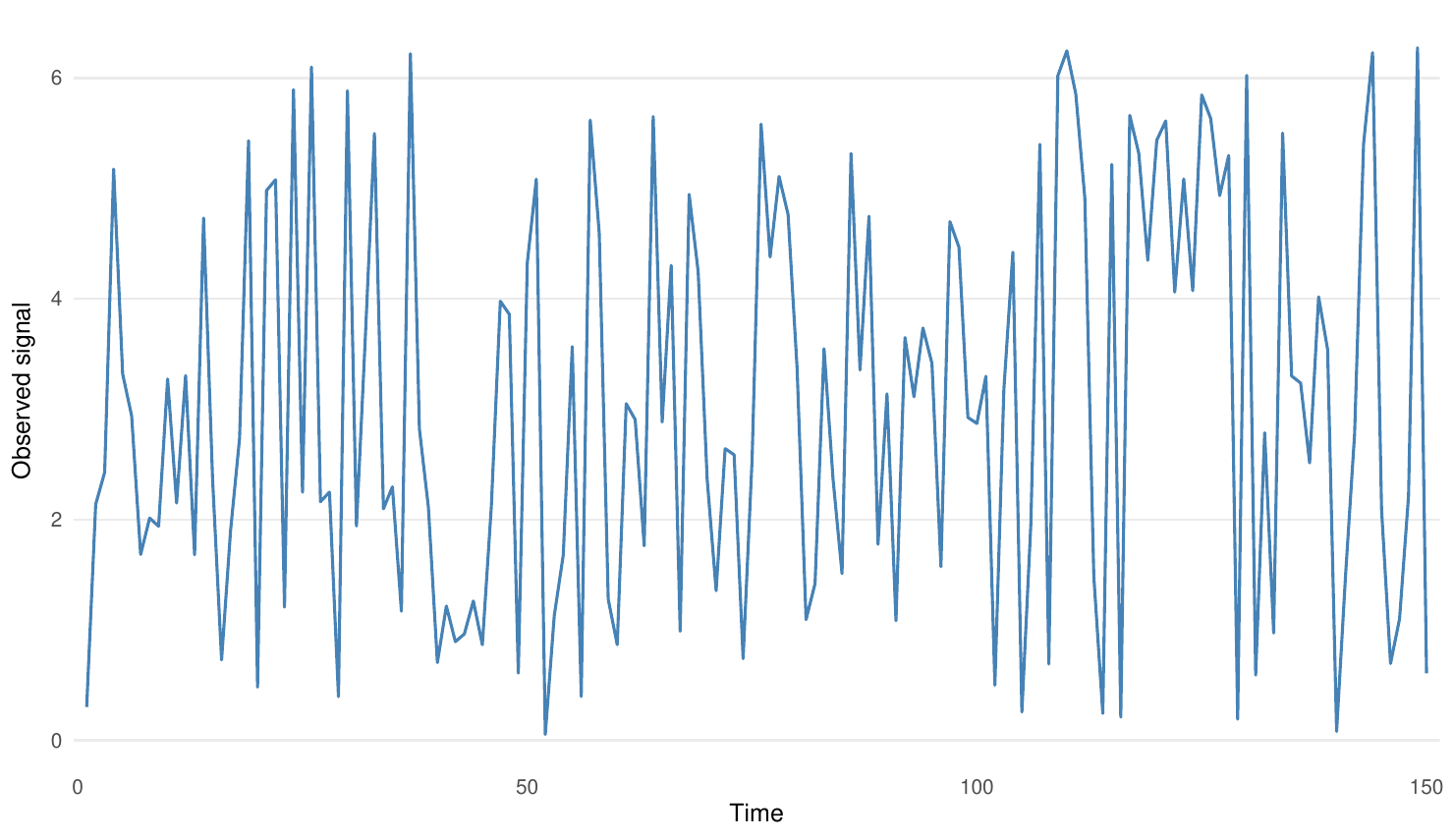} \hspace{0.7cm}
    \includegraphics[height = 5cm, width=7cm]{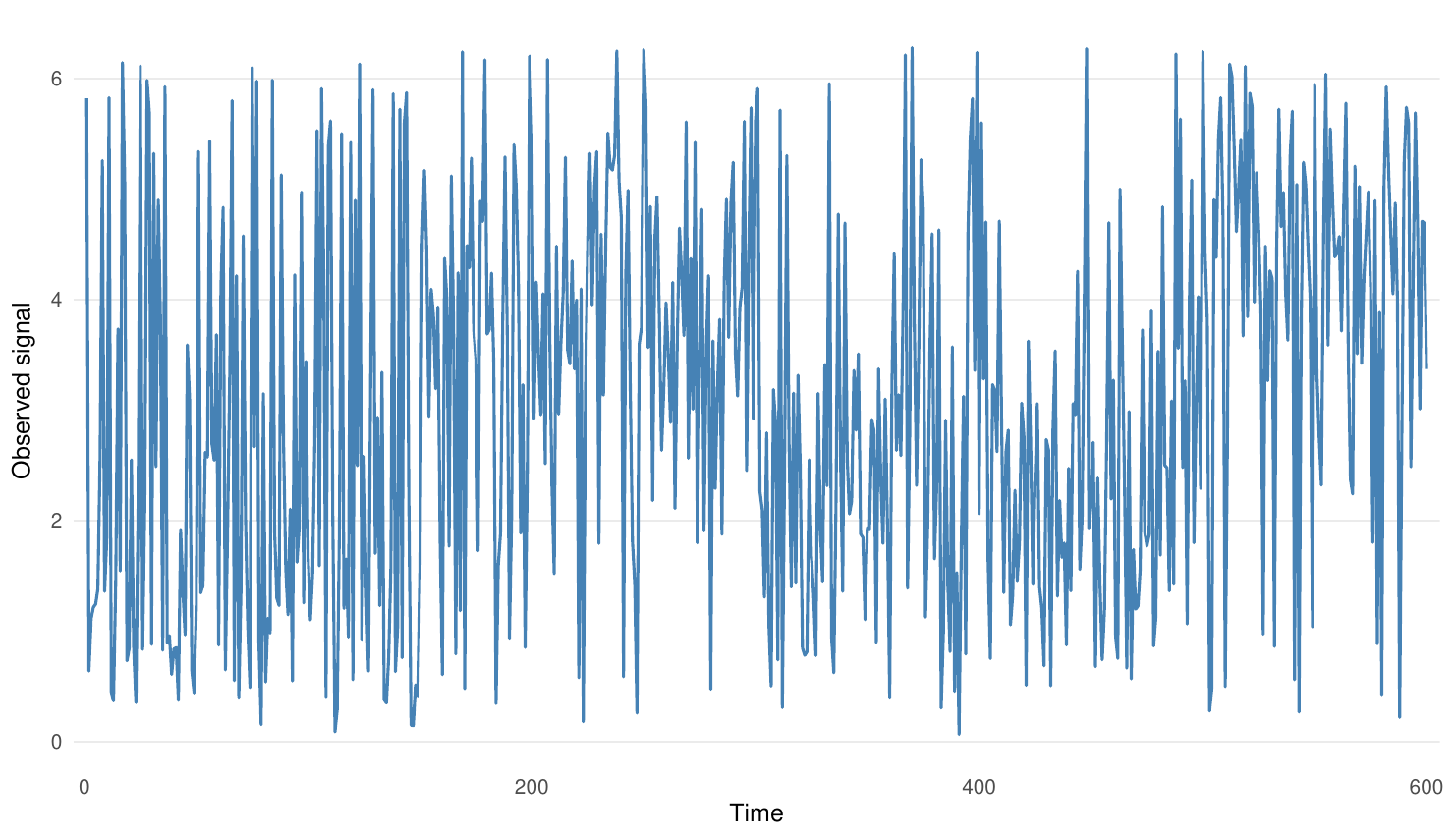} \\
		\hspace*{3.1cm} (a) \hspace{7.05cm} (b) \vspace{0.2cm}
    \caption{Plots of data sequences that follow model \eqref{eq: model} with $\epsilon_t \sim \text{vM}(0, \kappa)$ for (a) $\kappa = 2$ and (b) $\kappa = 1$ with underlying signals (a) \ref{signal: complex1} and (b) \ref{signal: long}.}
    \label{fig: signal_plot_complex_long}
\end{figure}

Some visualizations are given in Figures~\ref{fig: signal_plot_simple_simple2} and \ref{fig: signal_plot_complex_long}.
The plotted data sequences follow model \eqref{eq: model} when $\epsilon_t \sim \text{vM}(0, \kappa)$ for various values of $\kappa$, as provided in the caption of each plot.
The reader should keep in mind that the data have $2\pi$ periodicity when looking at the plots.
When the true mean of the data is close to $0$ or $2\pi$, due to the added noise, the changes can appear as change in the variance.
However, as the point $0$ can be chosen to be any point on the circle, a different choice of its location can change the way the plot looks.
It is worth noting that the change-points are not visible, yet the algorithm is able to accurately detect them (see Section~\ref{sec: independent_noise}).

\subsection{Independent noise structure}\label{sec: independent_noise}

\subsubsection{von Mises}\label{sec: von_mises}

In this subsection we consider model \eqref{eq: model} when $\epsilon_t \sim \text{vM}(0, \kappa)$, for $\kappa \in \{8,4,2,1\}$.
% It should be noted that since Signal \ref{signal: nocpt} contains no change-points, we only report the results for one choice of $\kappa$.
Our contrast function is constructed under the assumption that the data are von Mises distributed, so PCID is best suited for this type of noise.
In order to generate data we use the function \textit{rvonmises}() from the R package circular \citep{circular}.

The results can be found in Table~\ref{table: vonMises}.
As expected, setting $\gamma = 0.01$ makes the algorithm less prone to overestimation of change-points as compared to $\gamma = 0.05$.
When $\kappa = 1$, setting $\gamma = 0.01$ results in PCID  failing to detect some change-points, as the increased variance of the noise can mask the changes. 
This is especially evident for Signals~\ref{signal: simple3} and \ref{signal: complex1} whose magnitude of change at the change-points is smaller than the rest of the signals.
In such cases, the variance when $\kappa=1$ is larger than the magnitude of change, thus we observe under-detection of change-points even for $\gamma = 0.05$.
The simulation results also indicate that larger values of $\gamma$ can sometimes lead to higher computational times.
This can be explained by the fact that the algorithm needs to perform more permutations before deciding that there is no change-point (if, indeed, this is the case) in the interval under consideration (as $c_{B, \alpha_T}$ is larger).
Decreasing the concentration parameter also results in higher average computational times.
This is due to the fact that the algorithm struggles to identify the change-points (because of higher variability in the noise) and requires permutation testing to be done on more, hence larger intervals.

\begin{table}[htbp]
\caption{Distribution of $\hat{N} - N$ over 100 simulated data sequences of the Signals \ref{signal: nocpt} - \ref{signal: long} using the targeted Type I error mentioned, with $\epsilon_t \sim \text{vM}(0, \kappa)$. 
    The average ARI, $d_H$ and computational times are also given, when they are informative.} \label{table: vonMises}
\centering
        \begin{tabular}{|c|c|c|c|c|c|c|c|c|c|c|c|c|}
            \hline
            % & & & \multicolumn{7}{c|}{} & & & \\ 
            & & & \multicolumn{7}{c|}{$\hat{N} - N$} & & & \\
            Signal & Targeted $\gamma$ & $\kappa$ & $\leq -3$ & -2 & -1 & 0 & 1 & 2 & $\geq 3$ & $d_H$ & ARI & Time (s) \\
            \hline
            \multirow{2}{*}{\ref{signal: nocpt}} & 0.01 & 2 & - & - & - & 99 & 1 & 0 & 0 & - & - & 2.81 \\ 
            \cline{2-13}
            & 0.05 & 2 & - & - & - & 94 & 4 & 2 & 0 & - & - & 1.80 \\
            \hline
            \multirow{8}{*}{\ref{signal: simple}} & \multirow{4}{*}{0.01} & 8 & - & - & 0 & 97 & 3 & 0 & 0 & 0.010 & 0.995 & 0.97 \\
            & & 4 & - & - & 0 & 96 & 4 & 0 & 0 & 0.013 & 0.993 & 0.94 \\
            & & 2 & - & - & 0 & 98 & 2 & 0 & 0 & 0.012 & 0.982 & 1.02 \\
            & & 1 & - & - & 1 & 99 & 0 & 0 & 0 & 0.024 & 0.920 & 1.74 \\
            \cline{2-13}
            & \multirow{4}{*}{0.05} & 8 & - & - & 0 & 93 & 7 & 0 & 0 & 0.014 & 0.988 & 1.13 \\
            & & 4 & - & - & 0 & 91 & 9 & 0 & 0 & 0.028 & 0.982 & 1.12 \\
            & & 2 & - & - & 0 & 94 & 6 & 0 & 0 & 0.022 & 0.978 & 1.1 \\
            & & 1 & - & - & 0 & 98 & 1 & 1 & 0 & 0.025 & 0.923 & 1.67 \\
            \hline
            \multirow{8}{*}{\ref{signal: simple2}} & \multirow{4}{*}{0.01} & 8 & - & 0 & 0 & 97 & 3 & 0 & 0 & 0.010 & 0.995 & 0.97 \\
            & & 4 & - & 0 & 0 & 98 & 2 & 0 & 0 & 0.003 & 0.998 & 1.70 \\
            & & 2 & - & 0 & 0 & 98 & 2 & 0 & 0 & 0.007 & 0.989 & 1.94 \\
            & & 1 & - & 0 & 1 & 99 & 0 & 0 & 0 & 0.022 & 0.941 & 3.09 \\
            \cline{2-13}
            & \multirow{4}{*}{0.05} & 8 & - & 0 & 0 & 96 & 4 & 0 & 0 & 0.003 & 0.998 & 2.22 \\
            & & 4 & - & 0 & 0 & 96 & 4 & 0 & 0 & 0.005 & 0.994 & 2.11 \\
            & & 2 & - & 0 & 0 & 94 & 6 & 0 & 0 & 0.015 & 0.983 & 1.96 \\
            & & 1 & - & 0 & 1 & 95 & 4 & 0 & 0 & 0.028 & 0.938 & 2.95 \\
            \hline
            \multirow{8}{*}{\ref{signal: simple3}} & \multirow{4}{*}{0.01} & 8 & 0 & 0 & 0 & 100 & 0 & 0 & 0 & 0.002 & 1.000 & 2.84 \\
            & & 4 & 0 & 0 & 0 & 97 & 3 & 0 & 0 & 0.003 & 0.992 & 2.93 \\
            & & 2 & 0 & 0 & 0 & 98 & 2 & 0 & 0 & 0.011 & 0.964 & 3.93 \\
            & & 1 & 13 & 33 & 33 & 21 & 0 & 0 & 0 & 0.149 & 0.687 & 5.84 \\
            \cline{2-13}
            & \multirow{4}{*}{0.05} & 8 & 0 & 0 & 0 & 99 & 1 & 0 & 0 & 0.001 & 0.999 & 2.70 \\
            & & 4 & 0 & 0 & 0 & 95 & 5 & 0 & 0 & 0.004 & 0.991 & 2.72 \\
            & & 2 & 0 & 0 & 0 & 93 & 6 & 1 & 0 & 0.013 & 0.961 & 3.62 \\
            & & 1 & 4 & 23 & 41 & 31 & 1 & 0 & 0 & 0.115 & 0.740 & 5.66 \\
            \hline
            \multirow{8}{*}{\ref{signal: complex1}} & \multirow{4}{*}{0.01} & 8 & 0 & 0 & 0 & 98 & 2 & 0 & 0 & 0.001 & 0.998 & 1.51 \\
            & & 4 & 0 & 0 & 0 & 94 & 6 & 0 & 0 & 0.006 & 0.991 & 1.73 \\
            & & 2 & 0 & 0 & 0 & 96 & 4 & 0 & 0 & 0.015 & 0.972 & 2.22 \\
            & & 1 & 0 & 7 & 37 & 52 & 4 & 0 & 0 & 0.122 & 0.796 & 4.44 \\
            \cline{2-13}
            & \multirow{4}{*}{0.05} & 8 & 0 & 0 & 0 & 92 & 8 & 0 & 0 & 0.010 & 0.994 & 1.76 \\
            & & 4 & 0 & 0 & 0 & 92 & 8 & 0 & 0 & 0.011 & 0.987 & 1.64 \\
            & & 2 & 0 & 0 & 0 & 93 & 7 & 0 & 0 & 0.018 & 0.968 & 2.24 \\
            & & 1 & 0 & 1 & 26 & 69 & 4 & 0 & 0 & 0.081 & 0.84 & 4.10 \\
            \hline
            \multirow{8}{*}{\ref{signal: long}} & \multirow{4}{*}{0.01} & 8 & 0 & 0 & 0 & 93 & 7 & 0 & 0 & 0.004 & 0.998 & 10.9 \\
            & & 4 & 0 & 0 & 0 & 88 & 11 & 1 & 0 & 0.009 & 0.993 & 10.9 \\
            & & 2 & 0 & 0 & 0 & 89 & 7 & 4 & 0 & 0.012 & 0.990 & 11.1 \\
            & & 1 & 0 & 0 & 0 & 94 & 5 & 1 & 0 & 0.013 & 0.978 & 14.5 \\
            \cline{2-13}
            & \multirow{4}{*}{0.05} & 8 & 0 & 0 & 0 & 92 & 8 & 0 & 0 & 0.004 & 0.998 & 11.9 \\
            & & 4 & 0 & 0 & 0 & 87 & 11 & 2 & 0 & 0.010 & 0.993 & 10.9 \\
            & & 2 & 0 & 0 & 0 & 88 & 8 & 4 & 0 & 0.012 & 0.990 & 11.3 \\
            & & 1 & 0 & 0 & 0 & 92 & 6 & 2 & 0 & 0.015 & 0.977 & 13.8 \\
            \hline
\end{tabular}
\end{table}

\subsubsection{Different noise structures} \label{sec: different_noise}

We now consider different noise structures.
More specifically, we consider the wrapped Cauchy distribution, with probability density function given by 
\begin{equation*}
    f_{\text{wC}}(\theta; \mu, \rho) = \frac{1}{2\pi} \frac{1-\rho^2}{1+\rho^2-2\rho\cos(\theta-\mu)}
\end{equation*}
with concentration parameter $\rho \in (0,1)$,   and the wrapped Normal distribution with probability density function
\begin{equation*}
    f_{\text{wN}}(\theta; \mu, \beta) = \frac{1}{2\pi} \left( 1 + 2 \sum_{p=1}^\infty \beta^{p^2} \cos p(\theta - \mu) \right)
\end{equation*}
with concentration parameter $\beta \in [0,1]$.
% cardioid distribution with probability density function
% \begin{equation*}
%     f_c(\theta; \mu, \rho) = \frac{1}{2\pi} \left( 1 + 2 \beta \cos (\theta - \mu) \right)
% \end{equation*}
% with concentration parameter $\lvert \beta \rvert < \frac{1}{2}$.
In order to consider noise comparable with the von Mises distribution of Section~\ref{sec: von_mises}, we set the values of $\rho$ and $\beta$ such that the circular variances are equal to that of the von Mises setting.
The circular variance is given by $V = 1-R$, where $R$ is the population version of the mean resultant length as  in \eqref{eq: mean_resultant_length}.
It holds that
\begin{align*}
    V_{\text{vM}} & = 1 - \frac{I_1(\kappa)}{I_0(\kappa)},
    \ V_{\text{wC}} = 1 - \rho,
    \ V_{\text{wN}} = 1 - \beta,
\end{align*}
where $I_0(\kappa), I_1(\kappa)$ are given by \eqref{eq: bessel_function} and $V_{\text{vM}},V_{\text{wC}}, V_{\text{wN}}$ denote the circular variance for the von Mises, wrapped Cauchy and wrapped Normal distributions, respectively.
Numerical evaluation of the Bessel functions for $\kappa \in \{8,4,2,1\}$ yields corresponding values $\rho, \beta \in \{ 0.94, 0.86, 0.70, 0.45 \}$, respectively.
In order to generate data we use the functions \textit{rwrappedcauchy}() and \textit{rwrappednormal}() from the R package circular \citep{circular}.

In Table~\ref{table: wrpCauchy} we present the results of model \eqref{eq: model} for $\epsilon_t \sim \text{wC}(0, \rho)$, while in Table~\ref{table: wrpNormal} we consider $\epsilon_t \sim \text{wN}(0, \beta)$.
These data sequences do not follow the assumption under which our procedure is built.
However, as can be seen from the results, PCID performs well even in such scenarios.
This indicates that PCID seems robust under deviations from the von Mises distribution, upon which the contrast function was built.
% our contrast function is robust to the assumption of the underlying distribution of the noise and thus PCID is more generally applicable.

\begin{table}[tbp]
\caption{Distribution of $\hat{N} - N$ over 100 simulated data sequences of the Signals \ref{signal: nocpt} - \ref{signal: long} using the targeted Type I error mentioned, with $\epsilon_t \sim \text{wC}(0, \rho)$. 
    The average ARI, $d_H$ and computational times are also given, when they are informative.} \label{table: wrpCauchy}
\centering
        \begin{tabular}{|c|c|c|c|c|c|c|c|c|c|c|c|c|}
            \hline
            % & & & \multicolumn{7}{c|}{} & & & \\ 
            & & &\multicolumn{7}{c|}{$\hat{N} - N$} & & & \\
            Signal & Targeted $\gamma$ & $\rho$ & $\leq -3$ & -2 & -1 & 0 & 1 & 2 & $\geq 3$ & $d_H$ & ARI & Time (s) \\
            \hline
            \multirow{2}{*}{\ref{signal: nocpt}} & 0.01 & 0.86 & - & - & - & 95 & 2 & 3 & 0 & - & - & 1.58 \\ 
            \cline{2-13}
            & 0.05 & 0.86 & - & - & - & 98 & 2 & 0 & 0 & - & - & 0.83 \\
            \hline
            \multirow{8}{*}{\ref{signal: simple}} & \multirow{4}{*}{0.01} & 0.94 & - & - & 0 & 100 & 0 & 0 & 0 & 0.001 & 0.999 & 0.83 \\
            & & 0.86 & - & - & 0 & 99 & 1 & 0 & 0 & 0.006 & 0.991 & 0.84 \\
            & & 0.70 & - & - & 0 & 100 & 0 & 0 & 0 & 0.005 & 0.979 & 0.92 \\
            & & 0.45 & - & - & 2 & 95 & 3 & 0 & 0 & 0.032 & 0.914 & 1.47 \\
            \cline{2-13}
            & \multirow{4}{*}{0.05} & 0.94 & - & - & 0 & 100 & 0 & 0 & 0 & 0.001 & 0.999 & 0.98 \\
            & & 0.86 & - & - & 0 & 98 & 2 & 0 & 0 & 0.007 & 0.991 & 0.98 \\
            & & 0.70 & - & - & 0 & 95 & 5 & 0 & 0 & 0.023 & 0.975 & 0.98 \\
            & & 0.45 & - & - & 0 & 93 & 6 & 1 & 0 & 0.035 & 0.919 & 1.45 \\
            \hline
            \multirow{8}{*}{\ref{signal: simple2}} & \multirow{4}{*}{0.01} & 0.94 & - & 0 & 0 & 99 & 0 & 1 & 0 & 0.005 & 0.998 & 1.63 \\
            & & 0.86 & - & 0 & 0 & 98 & 1 & 1 & 0 & 0.009 & 0.992 & 1.65 \\
            & & 0.70 & - & 0 & 0 & 98 & 2 & 0 & 0 & 0.010 & 0.986 & 1.95 \\
            & & 0.45 & - & 1 & 0 & 98 & 1 & 0 & 0 & 0.019 & 0.943 & 2.74 \\
            \cline{2-13}
            & \multirow{4}{*}{0.05} & 0.94 & - & 0 & 0 & 97 & 2 & 1 & 0 & 0.006 & 0.990 & 1.84 \\
            & & 0.86 & - & 0 & 0 & 95 & 3 & 2 & 0 & 0.022 & 0.987 & 1.81 \\
            & & 0.70 & - & 0 & 0 & 90 & 7 & 3 & 0 & 0.027 & 0.977 & 2.06 \\
            & & 0.45 & - & 0 & 0 & 96 & 3 & 1 & 0 & 0.025 & 0.947 & 2.81 \\
            \hline
            \multirow{8}{*}{\ref{signal: simple3}} & \multirow{4}{*}{0.01} & 0.94 & 0 & 0 & 0 & 100 & 0 & 0 & 0 & 0.002 & 0.995 & 2.72 \\
            & & 0.86 & 0 & 0 & 0 & 97 & 3 & 0 & 0 & 0.005 & 0.987 & 2.95 \\
            & & 0.70 & 0 & 0 & 0 & 98 & 2 & 0 & 0 & 0.011 & 0.958 & 3.70 \\
            & & 0.45 & 14 & 26 & 38 & 22 & 0 & 0 & 0 & 0.137 & 0.711 & 5.83 \\
            \cline{2-13}
            & \multirow{4}{*}{0.05} & 0.94 & 0 & 0 & 0 & 98 & 2 & 0 & 0 & 0.004 & 0.994 & 2.69 \\
            & & 0.86 & 0 & 0 & 0 & 97 & 3 & 0 & 0 & 0.007 & 0.983 & 2.95 \\
            & & 0.70 & 0 & 0 & 0 & 97 & 3 & 0 & 0 & 0.012 & 0.961 & 3.56 \\
            & & 0.45 & 6 & 16 & 39 & 38 & 1 & 0 & 0 & 0.112 & 0.766 & 5.69 \\
            \hline
            \multirow{8}{*}{\ref{signal: complex1}} & \multirow{4}{*}{0.01} & 0.94 & 0 & 0 & 0 & 100 & 0 & 0 & 0 & 0.001 & 0.996 & 1.42 \\
            & & 0.86 & 0 & 0 & 0 & 99 & 1 & 0 & 0 & 0.007 & 0.991 & 1.51 \\
            & & 0.70 & 0 & 0 & 0 & 98 & 2 & 0 & 0 & 0.015 & 0.962 & 1.88 \\
            & & 0.45 & 1 & 14 & 26 & 58 & 1 & 0 & 0 & 0.144 & 0.780 & 3.38 \\
            \cline{2-13}
            & \multirow{4}{*}{0.05} & 0.94 & 0 & 0 & 0 & 94 & 6 & 0 & 0 & 0.005 & 0.992 & 1.54 \\
            & & 0.86 & 0 & 0 & 0 & 94 & 5 & 1 & 0 & 0.016 & 0.986 & 1.58 \\
            & & 0.70 & 0 & 0 & 0 & 96 & 4 & 0 & 0 & 0.017 & 0.961 & 1.94 \\
            & & 0.45 & 0 & 1 & 26 & 68 & 4 & 0 & 1 & 0.090 & 0.846 & 3.30 \\
            \hline
            \multirow{8}{*}{\ref{signal: long}} & \multirow{4}{*}{0.01} & 0.94 & 0 & 0 & 0 & 92 & 7 & 1 & 0 & 0.007 & 0.994 & 10.7 \\
            & & 0.86 & 0 & 0 & 0 & 96 & 2 & 2 & 0 & 0.007 & 0.995 & 10.8 \\
            & & 0.70 & 0 & 0 & 0 & 95 & 3 & 2 & 0 & 0.00928 & 0.990 & 12.2 \\
            & & 0.45 & 0 & 0 & 0 & 97 & 1 & 2 & 0 & 0.010 & 0.974 & 12.4 \\
            \cline{2-13}
            & \multirow{4}{*}{0.05} & 0.94 & 0 & 0 & 0 & 93 & 5 & 2 & 0 & 0.007 & 0.995 & 10.9 \\
            & & 0.86 & 0 & 0 & 0 & 93 & 3 & 3 & 1 & 0.005 & 0.993 & 11.0 \\
            & & 0.70 & 0 & 0 & 0 & 94 & 3 & 3 & 0 & 0.007 & 0.989 & 11.8 \\
            & & 0.45 & 0 & 0 & 0 & 91 & 8 & 1 & 0 & 0.015 & 0.967 & 13.4 \\
            \hline
\end{tabular}
\end{table}

\begin{table}[tbp]
\caption{Distribution of $\hat{N} - N$ over 100 simulated data sequences of the Signals \ref{signal: nocpt} - \ref{signal: long} using the targeted Type I error mentioned, with $\epsilon_t \sim \text{wN}(0, \beta)$. 
    The average ARI, $d_H$ and computational times are also given, when they are informative.} \label{table: wrpNormal}
\centering
        \begin{tabular}{|c|c|c|c|c|c|c|c|c|c|c|c|c|}
            \hline
            % & & & \multicolumn{7}{c|}{} & & & \\ 
            & & &\multicolumn{7}{c|}{$\hat{N} - N$} & & & \\
            Signal & Targeted $\gamma$ & $\beta$ & $\leq -3$ & -2 & -1 & 0 & 1 & 2 & $\geq 3$ & $d_H$ & ARI & Time (s) \\
            \hline
            \multirow{2}{*}{\ref{signal: nocpt}} & 0.01 & 0.86 & - & - & - & 97 & 3 & 0 & 0 & - & - & 0.67 \\ 
            \cline{2-13}
            & 0.05 & 0.86 & - & - & - & 96 & 1 & 3 & 0 & - & - & 1.35 \\
            \hline
            \multirow{8}{*}{\ref{signal: simple}} & \multirow{4}{*}{0.01} & 0.94 & - & - & 0 & 99 & 1 & 0 & 0 & 0.005 & 0.999 & 0.751 \\
            & & 0.86 & - & - & 0 & 100 & 0 & 0 & 0 & 0.001 & 1.00 & 0.756 \\
            & & 0.70 & - & - & 0 & 100 & 0 & 0 & 0 & 0.001 & 0.996 & 0.769 \\
            & & 0.45 & - & - & 0 & 97 & 3 & 0 & 0 & 0.020 & 0.951 & 1.25 \\
            \cline{2-13}
            & \multirow{4}{*}{0.05} & 0.94 & - & - & 0 & 93 & 6 & 1 & 0 & 0.018 & 0.987 & 0.858 \\
            & & 0.86 & - & - & 0 & 97 & 3 & 0 & 0 & 0.009 & 0.994 & 0.853 \\
            & & 0.70 & - & - & 0 & 93 & 6 & 1 & 0 & 0.026 & 0.983 & 0.887 \\
            & & 0.45 & - & - & 0 & 97 & 3 & 0 & 0 & 0.025 & 0.945 & 1.27 \\
            \hline
            \multirow{8}{*}{\ref{signal: simple2}} & \multirow{4}{*}{0.01} & 0.94 & - & 0 & 0 & 96 & 3 & 1 & 0 & 0.006 & 0.993 & 1.61 \\
            & & 0.86 & - & 0 & 0 & 95 & 5 & 0 & 0 & 0.014 & 0.991 & 1.65 \\
            & & 0.70 & - & 0 & 0 & 97 & 3 & 0 & 0 & 0.009 & 0.986 & 1.84 \\
            & & 0.45 & - & 0 & 0 & 98 & 2 & 0 & 0 & 0.015 & 0.957 & 2.60 \\
            \cline{2-13}
            & \multirow{4}{*}{0.05} & 0.94 & - & 0 & 0 & 90 & 6 & 4 & 0 & 0 0.0207 & 0.981 & 1.82 \\
            & & 0.86 & - & 0 & 0 & 97 & 2 & 1 & 0 & 0.0099 & 0.993 & 1.77 \\
            & & 0.70 & - & 0 & 0 & 95 & 5 & 0 & 0 & 0.0147 & 0.986 & 1.94 \\
            & & 0.45 & - & 0 & 0 & 94 & 6 & 0 & 0 & 0.0313 & 0.942 & 2.64 \\
            \hline
            \multirow{8}{*}{\ref{signal: simple3}} & \multirow{4}{*}{0.01} & 0.94 & 0 & 0 & 0 & 97 & 3 & 0 & 0 & 0.002 & 0.998 & 2.67 \\
            & & 0.86 & 0 & 0 & 0 & 97 & 3 & 0 & 0 & 0.003 & 0.992 & 2.78 \\
            & & 0.70 & 0 & 0 & 0 & 97 & 3 & 0 & 0 & 0.011 & 0.963 & 4.04 \\
            & & 0.45 & 14 & 30 & 37 & 19 & 0 & 0 & 0 & 0.132 & 0.702 & 7.23 \\
            \cline{2-13}
            & \multirow{4}{*}{0.05} & 0.94 & 0 & 0 & 0 & 95 & 3 & 2 & 0 & 0.002 & 0.998 & 2.72 \\
            & & 0.86 & 0 & 0 & 0 & 96 & 3 & 1 & 0 & 0.005 & 0.991 & 2.77 \\
            & & 0.70 & 0 & 0 & 0 & 94 & 6 & 0 & 0 & 0.013 & 0.961 & 3.56 \\
            & & 0.45 & 12 & 18 & 28 & 40 & 2 & 0 & 0 & 0.106 & 0.744 & 6.10 \\
            \hline
            \multirow{8}{*}{\ref{signal: complex1}} & \multirow{4}{*}{0.01} & 0.94 & 0 & 0 & 0 & 97 & 3 & 0 & 0 & 0.006 & 0.996 & 1.41 \\
            & & 0.86 & 0 & 0 & 0 & 97 & 3 & 0 & 0 & 0.005 & 0.992 & 1.49 \\
            & & 0.70 & 0 & 0 & 0 & 97 & 3 & 0 & 0 & 0.012 & 0.971 & 1.81 \\
            & & 0.45 & 1 & 8 & 34 & 56 & 1 & 0 & 0 & 0.130 & 0.799 & 3.31 \\
            \cline{2-13}
            & \multirow{4}{*}{0.05} & 0.94 & 0 & 0 & 0 & 94 & 3 & 3 & 0 & 0.011 & 0.992 & 1.55 \\
            & & 0.86 & 0 & 0 & 0 & 95 & 5 & 0 & 0 & 0.005 & 0.993 & 1.57 \\
            & & 0.70 & 0 & 0 & 97 & 2 & 1 & 0 & 0 & 0.011 & 0.970 & 1.82 \\
            & & 0.45 & 0 & 3 & 26 & 65 & 5 & 1 & 0 & 0.091 & 0.822 & 3.50 \\
            \hline
            \multirow{8}{*}{\ref{signal: long}} & \multirow{4}{*}{0.01} & 0.94 & 0 & 0 & 0 & 95 & 4 & 1 & 0 & 0.003 & 0.997 & 12.8 \\
            & & 0.86 & 0 & 0 & 0 & 92 & 5 & 3 & 0 & 0.006 & 0.995 & 13.4 \\
            & & 0.70 & 0 & 0 & 0 & 96 & 3 & 0 & 1 & 0.003 & 0.994 & 14.3 \\
            & & 0.45 & 0 & 0 & 0 & 94 & 5 & 1 & 0 & 0.012 & 0.975 & 14.4 \\
            \cline{2-13}
            & \multirow{4}{*}{0.05} & 0.94 & 0 & 0 & 0 & 90 & 7 & 2 & 1 & 0.008 & 0.992 & 11.3 \\
            & & 0.86 & 0 & 0 & 0 & 90 & 9 & 1 & 0 & 0.008 & 0.996 & 14.2 \\
            & & 0.70 & 0 & 0 & 0 & 94 & 5 & 1 & 0 & 0.007 & 0.991 & 14.6 \\
            & & 0.45 & 0 & 0 & 0 & 92 & 8 & 0 & 0 & 0.014 & 0.975 & 16.7 \\
            \hline
\end{tabular}
\end{table}

\subsection{Serially correlated data} \label{sec: correlated}
In this subsection we relax the assumption of independence of $\epsilon_t$.
In order to deal with such data, a common technique is to subsample the data sequence under consideration.
More specifically, the sequence $\Theta_1, \Theta_2, \ldots, \Theta_N$ is split into $\nu \in\N$ subsequences, denoted by $Y_{i;t}$ for $i \in \{1, 2, \ldots, \nu\}$ and $t \in \{ 1, \ldots, M_i \}$ where $M_i = \lfloor \frac{T-i}{\nu} \rfloor$, by choosing every $\nu^{\text{th}}$ observation, meaning that
\begin{equation*}
    \left \{ Y_{i;t} \right \}_{t=1}^{M_i} = \{ \Theta_i, \Theta_{i+\nu}, \ldots, \Theta_{i+M_i \nu } \}.
\end{equation*}
% \begin{equation*}
%     Y_{i;t} = \{ \Theta_i, \Theta_{i+\nu}, \ldots, \Theta_{M \nu + i} \}.
% \end{equation*}
PCID is applied to each of the $Y_{i;t}$, resulting in $\nu$ sequences of possible change-points.
A majority voting rule is applied to decide if the detected change-points in each of the subsequences should be regarded as change-points in the original one. 
For $\eta \in \{ 1, \ldots, \nu \}$, if at least $\eta$ sequences detect a change-point in the same position (or in a neighbourhood around it, say $\pm \delta$), we consider it to be a change-point.
If the change-point has been detected at the $j^{\text{th}}$ location of $Y_{i;t}$, its location in the original data sequence is $(j-1)\nu + i$.

In Table~\ref{table: dependent} we present the simulation results for Signals~\ref{signal: nocpt} and \ref{signal: simple}, when $\epsilon_t$ follows an autoregressive process with $p=1$ (AR(1))  with autoregressive coefficient $\phi \in \{0.3, 0.5, 0.7 \}$, given by $\epsilon_t = \phi \epsilon_{t-1} + \epsilon_t'$.
In order to respect the $2\pi$ periodic nature of circular data, the innovations, $\epsilon_t'$, are set to follow the von Mises distribution with concentration parameter $\kappa = 1.7$.
The usual choice for the innovations on the real line is $\epsilon_t' \sim N(0,1)$, which, when wrapped around the circle, has circular variance $V = 1-e^{-\frac{1}{2}}=0.4$ \citep[(3.5.63)]{mardia_directional_2000}.
For our choice of $\kappa$, it holds that $V_{\text{vM}} = 0.4$, where $V_{\text{vM}}$ denotes the circular variance of the von Mises distribution, meaning that we chose $\kappa$ such that the circular variances, $V$ and $V_{\text{VM}}$ coincide.
In our simulations we used $\nu = 5$, $\eta = 3$ and $\delta = 2$.

\begin{table}[tbp]
\caption{Distribution of $\hat{N} - N$ over 100 simulated data sequences of the Signals \ref{signal: nocpt}, \ref{signal: simple} using the targeted Type I error mentioned, with $\epsilon_t$ following AR(1) process with autoregressive coefficient $\phi$. 
    The average ARI, $d_H$ and computational times are also given, when they are informative.} \label{table: dependent}
\centering
        \begin{tabular}{|c|c|c|c|c|c|c|c|c|c|c|c|}
            \hline
            % & & & \multicolumn{5}{c|}{} & & & \\ 
            & & & \multicolumn{5}{c|}{$\hat{N} - N$} & & & \\
            Signal & Targeted Type I error & $\phi$ & -1 & 0 & 1 & 2 & $\geq 3$ & $d_H$ & ARI & Time (s) \\
            \hline
            \multirow{3}{*}{\ref{signal: nocpt}} & \multirow{3}{*}{0.05} & 0.3 & - & 99 & 1 & 0 & 0 & - & - & 0.130 \\ 
            & & 0.5 & - & 100 & 0 & 0 & 0 & - & - & 0.136 \\
            & & 0.7 & - & 91 & 8 & 1 & 0 & - & - & 0.172 \\
            \hline
            \multirow{3}{*}{\ref{signal: simple}} & \multirow{3}{*}{0.05} & 0.3 & 2 & 98 & 0 & 0 & 0 & 0.024 & 0.925 & 0.184 \\
            & & 0.5 & 18 & 82 & 0 & 0 & 0 & 0.109 & 0.750 & 0.179 \\
            & & 0.7 & 63 & 37 & 0 & 0 & 0 & 0.329 & 0.318 & 0.134 \\
            \hline
\end{tabular}
\end{table}

\section{Real data}\label{sec: real_data}

In this section we illustrate the practical relevance of PCID by applying it on three real world datasets.
In Section~\ref{sec: flare} we consider flare data, in Section~\ref{sec: acrophase} acrophase data and in Section~\ref{sec: wave} wave data.
While the first two examples have already been analysed for possible change-points in the literature \citep{Lombard1986, LOMBARD2017}, we are, to the best of our knowledge, the first to apply a change-point detection method on the third dataset.
% Our aim is to showcase the applicability of our proposed method to real world datasets, beyond the simulated data examples considered in Section~\ref{sec: simulations}.
Through these case studies we demonstrate the tangible impact of our method, besides its methodological advances as laid out in the Introduction.

\subsection{Flare data}\label{sec: flare}
\begin{figure}[tbp]
\centering
    \includegraphics[height = 6cm, width=8cm]{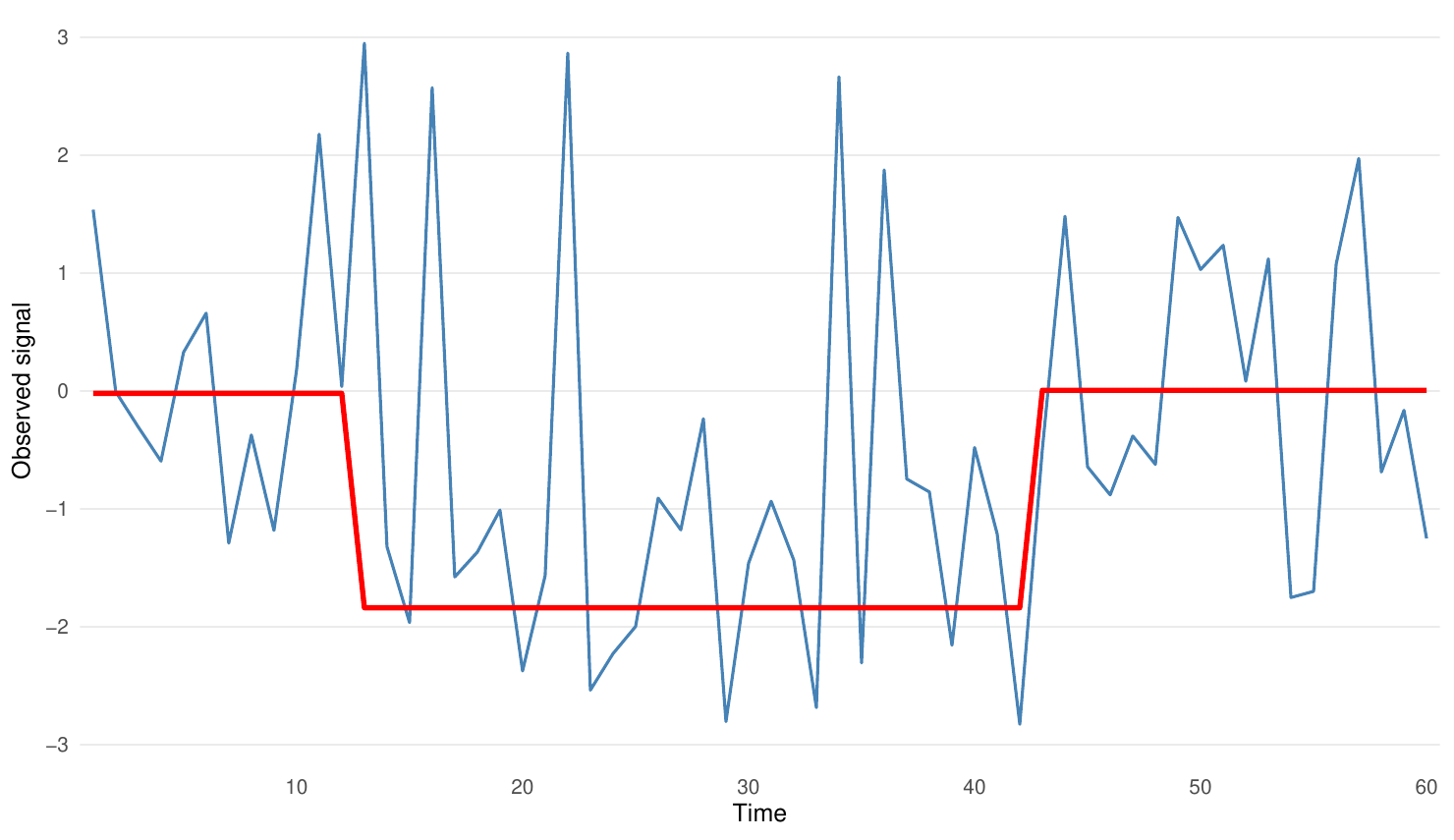}
    \caption{Plot of the flare data with the estimated signal from PCID in red.}
    \label{fig:flare_data}
\end{figure}

We firstly analyse data that were firstly considered in \cite{Lombard1986}.
The data come from a test system for illumination flares of the type commonly deployed in rescue operations. 
Each flare is fixed to a projectile that is fired from a point $O$ in a fixed direction, and begins burning at some point $P$ along its trajectory. The quantity of interest is the colatitude $\theta$ of the vector $OP$, whose dispersion characterizes the stability of the flare-projectile system. 
The dataset contains the colatitudes recorded for 60 consecutive launches of these assemblies and is plotted in Figure~\ref{fig:flare_data}.
For visualization purposes we plot the data on the $[-\pi, \pi)$ scale.

In our analysis we set $\gamma = 0.01$, for which $B=1000$ and $\alpha_T = 0.002$ are used.
The expansion parameter is set to $\lambda_T = 5$, as described in Section~\ref{sec: parameters}.
The change-points identified by PCID are at times 12 and 42.
These values coincide with the change-points detected by \cite{Lombard1986}.
% However, the change-points detected by \cite{LOMBARD2017} were 13 and 36.
The estimated signal is plotted in Figure~\ref{fig:flare_data}.

% \begin{figure}[tbp]
% 	\includegraphics[height = 5cm, width=7cm]{Plots/real_data_plots/Paper/flare_ggplot2-eps-converted-to.pdf}
%     \includegraphics[height = 5cm, width=7cm]{Plots/real_data_plots/Paper/flare_ggplot2_cpts-eps-converted-to.pdf} \\
% 		\hspace*{3.35cm} (a) \hspace{6.5cm} (b) \vspace{0.2cm}
%     \caption{(a) Plot of the flare data. (b) Plot of the flare data with the estimated signal from PCID in red.}
%     \label{fig:flare_data}
% \end{figure}

\subsection{Acrophase data}\label{sec: acrophase}
\begin{figure}[tbp]
\centering
    \includegraphics[height = 6cm, width=8cm]{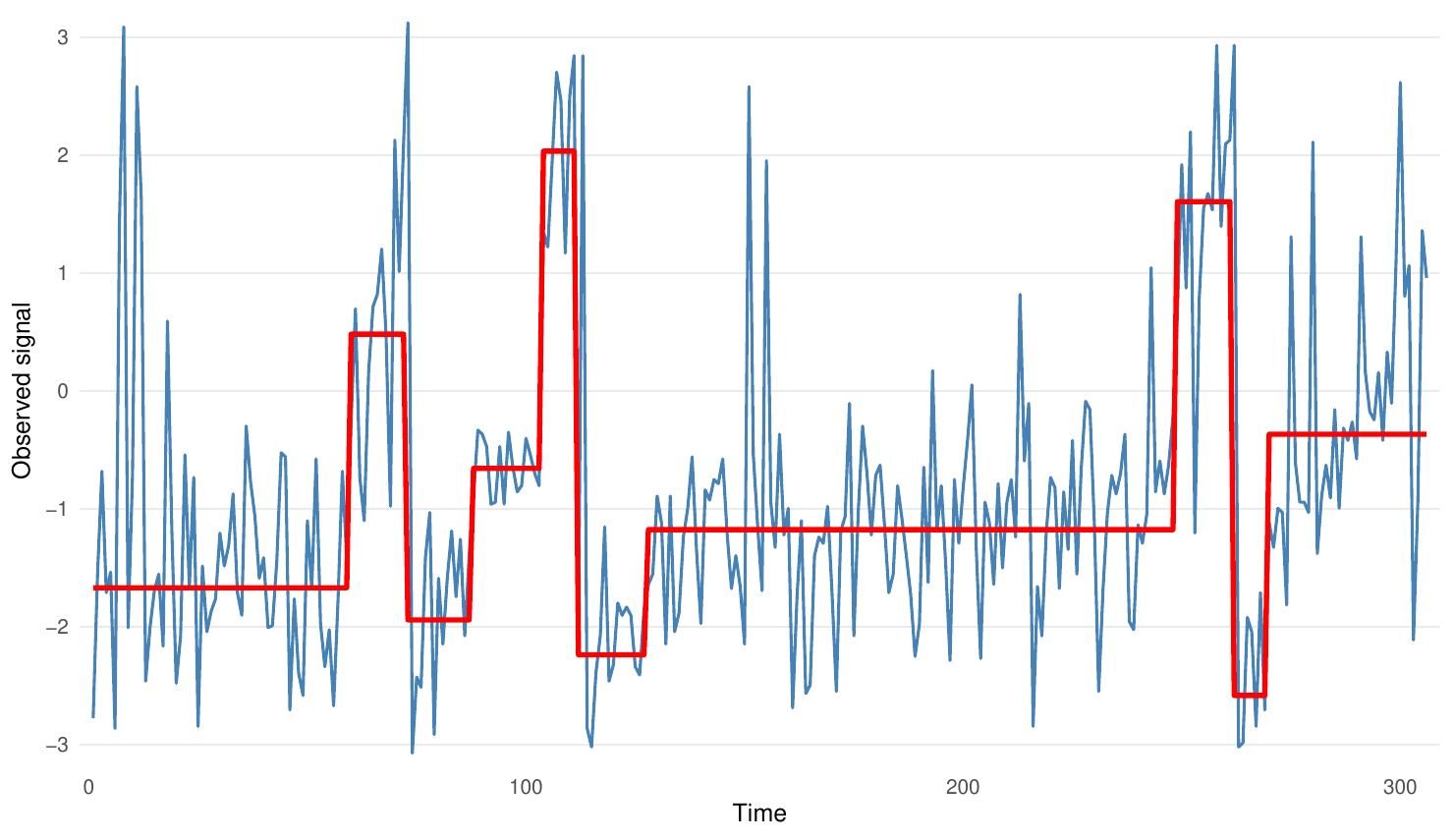}
    \caption{Plot of the acrophase data with the estimated signal from PCID in red.}
    \label{fig:acrophase_data}
\end{figure}

We now analyse data that are also considered in \cite{LOMBARD2017}.
The scope of \cite{LOMBARD2017} is online change-point detection, thus we do not compare our results with their detected change-points.
Acrophase is the time at which the maximum systolic blood pressure is reached on a given day. 
Monitoring such data for changes in the time that the maximum is achieved in different days can provide an early warning sign of a possible medical condition before it becomes clinically obvious.
The data was collected from a patient suffering from episodes of clinical depression.
The dataset contains 306 observations and is plotted in Figure~\ref{fig:acrophase_data}.
For visualization purposes we plot the data on the $[-\pi, \pi)$ scale.

In our analysis we set $\gamma = 0.01$, for which $B=1000$ and $\alpha_T = 0.001$ are used, and $\lambda_T = 5$. Our algorithm detected 9 change-points at times $59,  72,  87, 103, 111, 127, 248, 261, 269$. 
% We do not compare our results with the ones obtained by \cite{LOMBARD2017}, as their algorithm concerns online change-point detection.
The estimated signal from PCID are plotted in Figure~\ref{fig:acrophase_data}.

% \begin{figure}[tbp]
% 	\includegraphics[height = 5cm, width=7cm]{Plots/real_data_plots/Paper/acrophase_ggplot2-eps-converted-to.pdf}
%     \includegraphics[height = 5cm, width=7cm]{Plots/real_data_plots/Paper/acrophase_ggplot2_cpts-eps-converted-to.pdf} \\
% 		\hspace*{3.35cm} (a) \hspace{6.5cm} (b) \vspace{0.2cm}
%     \caption{(a) Plot of the acrophase data. (b) Plot of the acrophase data with the estimated signal from PCID in red.}
%     \label{fig:acrophase_data}
% \end{figure}

\subsection{Wave data}\label{sec: wave}
\begin{figure}[tbp]
    \centering
    \includegraphics[height = 6cm, width=8cm]{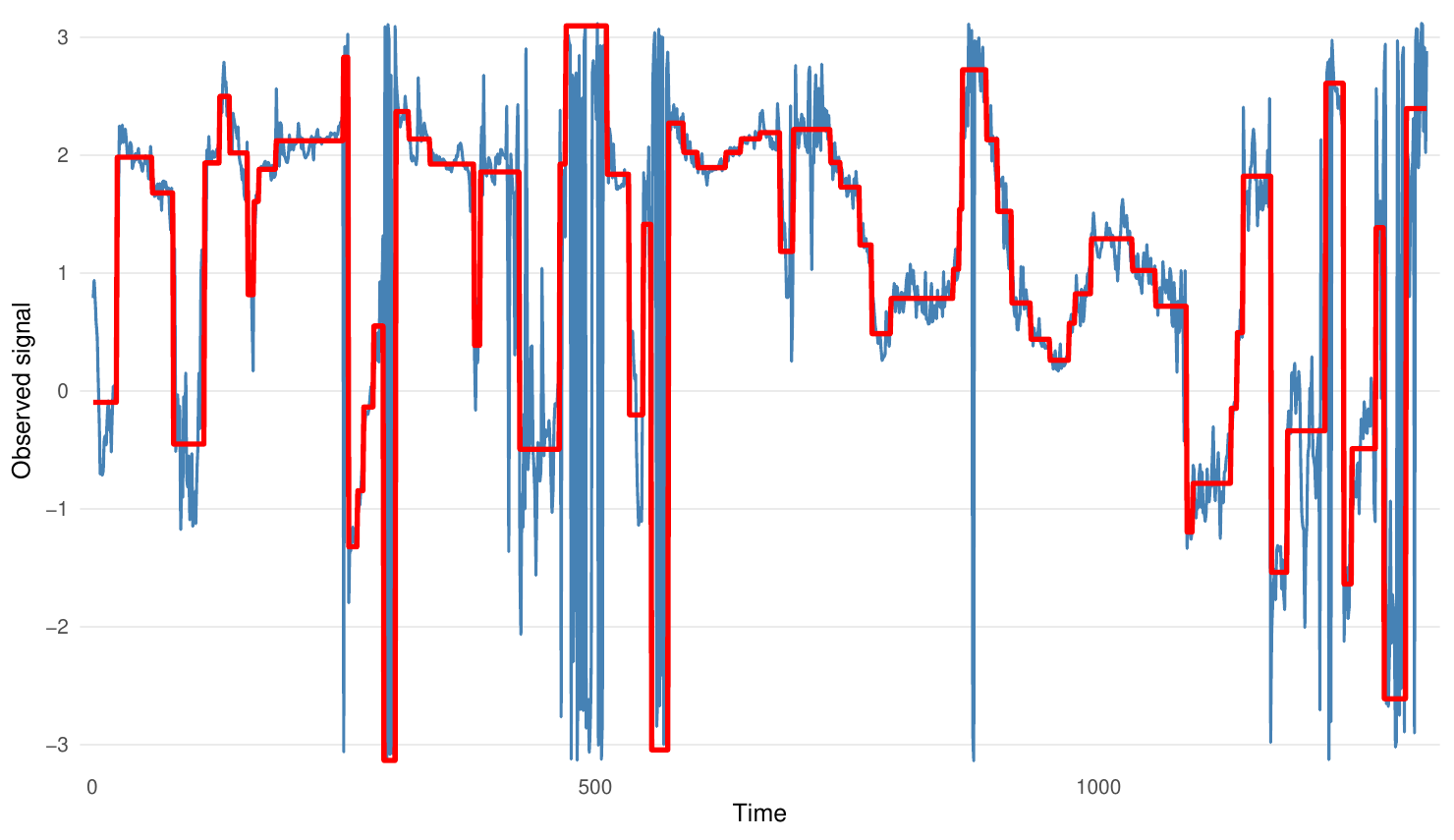}
    \caption{Plot of the wave data with the estimated signal from PCID in red.}
    \label{fig:wave_data}
\end{figure}

We now consider a dataset containing wave directions, recorded in the period 15/02/2010 – 16/03/2010 by the buoy of Ancona, located in the Adriatic Sea at about 30 km from the coast.
The data consists of 1326 observations of half-hourly wave directions.
The data sequence is plotted in Figure~\ref{fig:wave_data}.
For visualization purposes we plot the data on the $[-\pi, \pi)$ scale.
One could mistakenly conclude that there is a big variability between observations, especially around $t=400-600$.
However, due to periodicity, the values close to $-\pi$ and $\pi$ are in fact close to each other.
This dataset was also analysed by \cite{lagona_hidden_2015} and \cite{loizidou2025modellingtoroidalcylindricaldata}, but not for the purpose of change-point detection.

The data sequence has length $T = 1326$, thus, in order to reduce the computational complexity of our algorithm, we employ the PCID$_\text{W}$ variant with $w=500$, as described in Section~\ref{sec: variants}.
This implies that the sequence is split into 3 disjoint intervals and, since
we set $\gamma=0.01$, $\gamma_i = 0.003$ is used for $i=1,2,3$, where $\gamma_i$ is defined in \eqref{eq: gamma_i}.
For all intervals under consideration, the parameter values used are $B = 1000$, $\alpha_T = 0.001$ and $\lambda_T = 5$.
PCID detects 68 change-points.
The estimated signal is plotted in Figure~\ref{fig:wave_data}.
We also carried out a more conservative analysis of the data, by forcing the algorithm to use $B = 10,000$ and the appropriate values from Table~\ref{tab: alpha} for each disjoint interval.
With this analysis we obtained 63 change-points, so the results are very similar.
% If a more conservative analysis of the dataset was performed, we could force the algorithm to use $B = 10,000$ and the appropriate values from Table~\ref{tab: alpha} for each disjoint interval.
% This dataset has not been analysed for change-points before, so we do not compare with any competitors.
% \begin{figure}[tbp]
%     \centering
%     \includegraphics[height = 5cm, width=7cm]{Plots/real_data_plots/Paper/wave_ggplot2-eps-converted-to.pdf}
%     \includegraphics[height = 5cm, width=7cm]{Plots/real_data_plots/Paper/wave_ggplot2_cpts-eps-converted-to.pdf}\\
% 		\hspace*{0.45cm} (a) \hspace{6.45cm} (b) \vspace{0.2cm}
%     \caption{(a) Plot of the wave data. (b) Plot of the wave data with the estimated signal from PCID in red.}
%     \label{fig:wave_data}
% \end{figure}

\section{Possible extensions/discussion} \label{sec: conclusions}

In this paper we have proposed a new change-point detection algorithm for changes in the mean of circular data, denoted PCID.
The algorithm first isolates each change-point and then detects it using permutation testing. As highlighted in Sections~\ref{sec: intro} and \ref{sec: methodology}, the isolation aspect of the method is very important as detection becomes easier when a single change-point is present in the interval under consideration.
% increases the detection power of PCID.
For the detection of change-points, a contrast function is employed, which is constructed under the assumption that the noise, $\epsilon_t$ of model \eqref{eq: model}, follows the von Mises distribution.
Through simulations we show that the algorithm can also be applied to data sequences that do not satisfy this distributional assumption.
The performance of the algorithm is also investigated for serially correlated noise structures.

A different, not very explored area of change-point detection, concerns multidimensional manifolds involving angles, such as the torus or cylinder, which are the Cartesian product of circles ($[0,2\pi)^2$) and the Cartesian product between a circle and a real-valued variable ($[0,2\pi)\times\R$), respectively.
The wave dataset considered in Section~\ref{sec: wave} is part of a bigger dataset which also involves wind directions and wave height \citep{loizidou2025modellingtoroidalcylindricaldata}.
Toroidal data can arise when considering wave and wind directions, while wave direction and height make up a cylindrical dataset.
A natural direction for future research would be to extend PCID to toroidal or cylindrical data.

%%%%%%%%%%%%%%

\begin{appendix}

\section{Simulations for different expansion parameters} \label{sec: lambda_sims}
In this section we present some simulation results for different choices of the expansion parameter, $\lambda_T$. 
More specifically we consider $\lambda_T \in \{2,3,4,5,10,20,30,40,50\}$.
The signals used are the following
\begin{itemize}
    \item[\textlabel{(S9)}{signal: nocpt_long}]  sequence of length $T=500$ with no change-points. 

    \item[\textlabel{(S10)}{signal: teeth}]  sequence of length $T=500$ with 9 change-points at locations $50, 100, \ldots, 450$ and values between change-points $0, 1.5, 0, \ldots, 0, 1.5$.

    \item[\textlabel{(S11)}{signal: teeth_small}]  sequence of length $T=100$ with 9 change-points at locations $10, 20, \ldots, 90$ and values between change-points $0, 2, 0, \ldots, 0, 2$.
\end{itemize}
For Signals \ref{signal: nocpt_long} and \ref{signal: teeth}, we use $\alpha_T = 0.001$, while for Signal \ref{signal: teeth_small} we use $\alpha_T = 0.003$. 
For all three signals we set $B = 1000$ and $\epsilon_t \sim \text{vM}(0, 4)$.
The values chosen for $\alpha_T$ and $B$ are in line with Section~\ref{sec: parameters}.
We work under the same framework as described in Section~\ref{sec: framework}.
Table~\ref{table: sims_lambda} indicates that in cases of no change-points as in Signal \ref{signal: nocpt_long}, the average computational time decreases as $\lambda_T$ increases.
When change-points are present, the conclusions are different.
The simulation results concerning Signal~\ref{signal: teeth} show that the computational time is not directly proportional to the value of $\lambda_T$.
When the change-points are close to each other, as in Signal~\ref{signal: teeth_small}, large values of $\lambda_T$ fail to detect any change-points.

\begin{table}[tbp]
\caption{Distribution of $\hat{N} - N$ over 100 simulated data sequences of the Signals \ref{signal: nocpt_long}, \ref{signal: teeth} and \ref{signal: teeth_small} for $\epsilon_t \sim \text{vM}(0, 4)$. 
    The average ARI, $d_H$ and computational times are also given, when they are informative.} \label{table: sims_lambda}
\centering
        \begin{tabular}{|c|c|c|c|c|c|c|c|c|c|c|c|}
            \hline
            % & & \multicolumn{7}{c|}{} & & & \\ 
            & & \multicolumn{7}{c|}{$\hat{N} - N$} & & & \\
            Signal & $\lambda_T$ & $\leq -3$ & -2 & -1 & 0 & 1 & 2 & $\geq 3$ & $d_H$ & ARI & Time (s) \\
            \hline
            \multirow{9}{*}{\ref{signal: nocpt_long}} & 2 & - & - & - & 88 & 5 & 6 & 1 & - & - & 16.8 \\ 
            & 3 & - & - & - & 90 & 4 & 5 & 1 & - & - & 12.6 \\
            & 4 & - & - & - & 91 & 5 & 3 & 1 & - & - & 9.06 \\
            & 5 & - & - & - & 93 & 3 & 4 & 0 & - & - & 8.29 \\
            & 10 & - & - & - & 96 & 2 & 2 & 0 & - & - & 3.50 \\
            & 20 & - & - & - & 97 & 2 & 1 & 0 & - & - & 1.86 \\
            & 30 & - & - & - & 98 & 1 & 1 & 0 & - & - & 1.46 \\
            & 40 & - & - & - & 98 & 1 & 1 & 0 & - & - & 1.24 \\
            & 50 & - & - & - & 97 & 2 & 1 & 0 & - & - & 1.04 \\
            \hline
            \multirow{9}{*}{\ref{signal: teeth}} & 2 & - & 0 & 0 & 85 & 15 & 0 & 0 & 0.006 & 0.996 & 12.2 \\ 
            & 3 & - & 0 & 0 & 90 & 10 & 0 & 0 & 0.004 & 0.997 & 9.71 \\
            & 4 & - & 0 & 0 & 90 & 10 & 0 & 0 & 0.003 & 0.998 & 12.7 \\
            & 5 & - & 0 & 0 & 97 & 3 & 0 & 0 & 0.002 & 0.999 & 8.97 \\
            & 10 & - & 0 & 0 & 94 & 6 & 0 & 0 & 0.002 & 0.998 & 9.42 \\
            & 20 & - & 0 & 0 & 95 & 5 & 0 & 0 & 0.002 & 0.999 & 9.32 \\
            & 30 & - & 0 & 0 & 97 & 3 & 0 & 0 & 0.002 & 0.999 & 9.27 \\
            & 40 & - & 0 & 0 & 99 & 1 & 0 & 0 & 0.001 & 0.999 & 13.2 \\
            & 50 & - & 0 & 0 & 99 & 1 & 0 & 0 & 0.001 & 0.999 & 16.7 \\
            \hline
            \multirow{9}{*}{\ref{signal: teeth_small}} & 2 & 0 & 1 & 2 & 93 & 4 & 0 & 0 & 0.013 & 0.971 & 3.17 \\
            & 3 & 0 & 1 & 1 & 97 & 1 & 0 & 0 & 0.010 & 0.973 & 2.80 \\
            & 4 & 0 & 0 & 2 & 96 & 2 & 0 & 0 & 0.011 & 0.974 & 2.71 \\
            & 5 & 0 & 0 & 3 & 95 & 2 & 0 & 0 & 0.011 & 0.974 & 2.36 \\
            & 10 & 0 & 0 & 2 & 96 & 2 & 0 & 0 & 0.010 & 0.977 & 2.71 \\
            & 20 & 0 & 1 & 0 & 98 & 1 & 0 & 0 & 0.008 & 0.977 & 2.60 \\
            & 30 & 100 & 0 & 0 & 0 & 0 & 0 & 0 & 0.824 & 0.030 & 1.03 \\
            & 40 & 100 & 0 & 0 & 0 & 0 & 0 & 0 & 0.703 & 0.102 & 1.32 \\
            & 50 & 100 & 0 & 0 & 0 & 0 & 0 & 0 & 0.891 & 0.004 & 0.60  \\
            \hline
\end{tabular}
\end{table}

\end{appendix}

%%%%%%%%%%%%%%%%%%%%%%%%%%%%%%%%%%%%%%%%%%%%%%
%% Single Appendix:                         %%
%%%%%%%%%%%%%%%%%%%%%%%%%%%%%%%%%%%%%%%%%%%%%%
%\begin{appendix}
%\section*{???}%% if no title is needed, leave empty \section*{}.
%\end{appendix}
%%%%%%%%%%%%%%%%%%%%%%%%%%%%%%%%%%%%%%%%%%%%%%
%% Multiple Appendixes:                     %%
%%%%%%%%%%%%%%%%%%%%%%%%%%%%%%%%%%%%%%%%%%%%%%
% \begin{appendix}

% \end{appendix}

%%%%%%%%%%%%%%%%%%%%%%%%%%%%%%%%%%%%%%%%%%%%%%
%% Support information, if any,             %%
%% should be provided in the                %%
%% Acknowledgements section.                %%
%%%%%%%%%%%%%%%%%%%%%%%%%%%%%%%%%%%%%%%%%%%%%%

\

\textbf{Code availability:}
The code is available in \url{https://github.com/Sophia-Loizidou/PCID}.

\

\textbf{Data availability:}
The flare and acrophase datasets are available in the online Supplementary Material of \cite{LOMBARD2017}.
The wave dataset is available upon request to S.L.

\

\textbf{Acknowledgments: }
The authors would like to thank Francesco Lagona for the wave data.
S.L. was funded by the Luxembourg National Research Fund (FNR), grant reference PRIDE/21/16747448/MATHCODA.

% \begin{supplement}
% \stitle{Supplement to `Construction of optimal tests for symmetry on the torus and their quantitative error bounds'}
% \sdescription{The online Supplementary Material contains the proofs of all propositions, theorems and lemmas as well as statements and proofs of some useful lemmas for the proofs provided.
% It also includes some further simulation results.}
% \end{supplement}

% \bibliographystyle{unsrtnat}
\bibliographystyle{abbrvnat}
\bibliography{references}  %%% Uncomment this line and comment out the ``thebibliography'' section below to use the external .bib file (using bibtex) .

@book{mardia_directional_2000,
	address = {Chichester, United Kingdom},
	title = {Directional {Statistics}},
	publisher = {John Wiley \& Sons},
	author = {Mardia, Kanti V and Jupp, Peter E},
	year = {2000},
}

@article{ghosh_change-point_1999,
	title = {Change-point problems for the von {Mises} distribution},
	volume = {26},
	number = {4},
	urldate = {2024-09-23},
	journal = {Journal of Applied Statistics},
	author = {Ghosh, Kaushik and Jammalamadaka, S. Rao and Vasudaven, Mangalam},
	year = {1999},
	publisher = {Taylor \& Francis},
	pages = {423--434},
}

@article{grabovsky_change-point_2001,
	title = {Change-{Point} {Detection} in {Angular} {Data}},
	volume = {53},
	language = {en},
	number = {3},
	urldate = {2025-11-24},
	journal = {Annals of the Institute of Statistical Mathematics},
	author = {Grabovsky, Irina and Horváth, Lajos},
	year = {2001},
	pages = {552--566},
}

@misc{xu_change_2025,
	title = {Change {Point} {Detection} for {Random} {Objects} with {Periodic} {Behavior}},
	urldate = {2025-11-24},
	publisher = {arXiv},
	author = {Xu, Jiazhen and Wood, Andrew T. A. and Zou, Tao},
	year = {2025},
	note = {arXiv:2501.01657 [stat]},
}

@article{sengupta_bayesian_2008,
	title = {A {Bayesian} analysis of the change-point problem for directional data},
	volume = {35},
	number = {6},
	urldate = {2025-11-24},
	journal = {Journal of Applied Statistics},
	author = {Sengupta, Ashis and Kumar Laha, Arnab},
	year = {2008},
	publisher = {Taylor \& Francis},
	pages = {693--700},
}

@article{CSORGOXB199661,
title = {A note on the change-point problem for angular data},
journal = {Statistics \& Probability Letters},
volume = {27},
number = {1},
pages = {61-65},
year = {1996},
author = {Cs{\"o}rgő, Mikl{\'o}s and Horv{\'a}th, Lajos}
}

@article{Lombard1986,
 author = {F. Lombard},
 journal = {Technometrics},
 number = {4},
 pages = {391--397},
 publisher = {[Taylor \& Francis, Ltd., American Statistical Association, American Society for Quality]},
 title = {The Change-Point Problem for Angular Data: A Nonparametric Approach},
 urldate = {2025-11-25},
 volume = {28},
 year = {1986}
}

@Inbook{Potgieter2022,
author="Potgieter, Cornelis J.
and Lombard, F.
and Hawkins, Douglas M.",
editor="SenGupta, Ashis
and Arnold, Barry C.",
title="Statistical Process Control on the Circle: A Review and Some New Results",
bookTitle="Directional Statistics for Innovative Applications: A Bicentennial Tribute to Florence Nightingale",
year="2022",
publisher="Springer Nature Singapore",
address="Singapore",
pages="425--447",
isbn="978-981-19-1044-9",
}

@article{Lacomba03052025,
author = {Diego Lacomba and Amelia Simó and Oscar Belmonte},
title = {Detection of anomalous behaviour using mixtures of von {Mises} distributions and an extension of the {CUSUM} algorithm},
journal = {Journal of Statistical Computation and Simulation},
volume = {95},
number = {7},
pages = {1556--1571},
year = {2025},
publisher = {Taylor \& Francis},
}

@article{hubert1985comparing,
  title={Comparing partitions},
  author={Hubert, Lawrence and Arabie, Phipps},
  journal={Journal of Classification},
  volume={2},
  number={1},
  pages={193--218},
  year={1985},
  publisher={Springer}
}

@article{LOMBARD2017,
title = {Sequential rank {CUSUM} charts for angular data},
journal = {Computational Statistics \& Data Analysis},
volume = {105},
pages = {268-279},
year = {2017},
author = {F. Lombard and Douglas M. Hawkins and Cornelis J. Potgieter},
}

@article{lombaard1991changepoint,
  title={A changepoint analysis of some data arising in gamma-ray astronomy},
  author={Lombard, F},
  journal={South African Statistical Journal},
  volume={25},
  number={2},
  pages={83--98},
  year={1991},
  publisher={South African Statistical Association (SASA)}
}

@article{LOMBARD1990285,
title = {The detection of a change point in periodic gamma ray data},
journal = {Nuclear Physics B - Proceedings Supplements},
volume = {14},
number = {1},
pages = {285-290},
year = {1990},
author = {F. Lombard and O.C. {de Jager} and D.M. Schultz},
}

@article{ANTOCH200137,
title = {Permutation tests in change point analysis},
journal = {Statistics \& Probability Letters},
volume = {53},
number = {1},
pages = {37-46},
year = {2001},
author = {Jaromı́r Antoch and Marie Hušková},
}

@article{lombard2012cusum,
  title={A cusum procedure to detect deviations from uniformity in angular data},
  author={Lombard, F and Maxwell, RK},
  journal={Journal of Applied Statistics},
  volume={39},
  number={9},
  pages={1871--1880},
  year={2012},
  publisher={Taylor \& Francis}
}

@article{Gadsden1981,
 author = {R. J. Gadsden and G. K. Kanji},
 journal = {Journal of the Royal Statistical Society. Series D (The Statistician)},
 number = {2},
 pages = {119--129},
 publisher = {[Royal Statistical Society, Wiley]},
 title = {Sequential Analysis for Angular Data},
 urldate = {2025-12-12},
 volume = {30},
 year = {1981}
}

@article{Hawkins11062017,
author = {Douglas M. Hawkins and F. Lombard},
title = {Cusum control for data following the von {Mises} distribution},
journal = {Journal of Applied Statistics},
volume = {44},
number = {8},
pages = {1319--1332},
year = {2017},
publisher = {Taylor \& Francis},
}

@article{huvskova2001permutation,
  title={Permutation tests for multiple changes},
  author={Hu{\v{s}}kov{\'a}, Marie and Slab{\`y}, Ale{\v{s}}},
  journal={Kybernetika},
  volume={37},
  number={5},
  pages={605--622},
  year={2001},
  publisher={Institute of Information Theory and Automation AS CR}
}

@article{zeileis2013toolbox,
  title={A toolbox of permutation tests for structural change},
  author={Zeileis, Achim and Hothorn, Torsten},
  journal={Statistical Papers},
  volume={54},
  number={4},
  pages={931--954},
  year={2013},
  publisher={Springer}
}

@article{HORVATH2005351,
title = {Testing for changes using permutations of {U}-statistics},
journal = {Journal of Statistical Planning and Inference},
volume = {128},
number = {2},
pages = {351-371},
year = {2005},
author = {Lajos Horváth and Marie Hušková},
}

@article{berry2014chronicle,
  title={A chronicle of permutation statistical methods},
  author={Berry, Kenneth J and Johnston, Janis E and Mielke Jr, Paul W},
  journal={Cham: Springer},
  year={2014},
  publisher={Springer}
}

@article{huh2001random,
  title={Random permutation testing in multiple linear regression},
  author={Huh, Myung-Hoe and Jhun, Myoungshic},
  journal={Communications in Statistics-Theory and Methods},
  volume={30},
  number={10},
  pages={2023--2032},
  year={2001},
  publisher={Taylor \& Francis}
}

@article{long2009tetrachoric,
  title={Tetrachoric correlation: A permutation alternative},
  author={Long, Michael A and Berry, Kenneth J and Mielke Jr, Paul W},
  journal={Educational and Psychological Measurement},
  volume={69},
  number={3},
  pages={429--437},
  year={2009},
  publisher={Sage Publications Sage CA: Los Angeles, CA}
}

@article{reiss2010distance,
  title={On distance-based permutation tests for between-group comparisons},
  author={Reiss, Philip T and Stevens, M Henry H and Shehzad, Zarrar and Petkova, Eva and Milham, Michael P},
  journal={Biometrics},
  volume={66},
  number={2},
  pages={636--643},
  year={2010},
  publisher={Oxford University Press}
}

@article{jung2007new,
  title={A new random permutation test in {ANOVA} models},
  author={Jung, Byoung Cheol and Jhun, Myoungshic and Song, Seuck Heun},
  journal={Statistical Papers},
  volume={48},
  number={1},
  pages={47--62},
  year={2007},
  publisher={Springer}
}

@article{ernst2004permutation,
 title = {Permutation Methods: A Basis for Exact Inference},
  author = {Michael D. Ernst},
 journal = {Statistical Science},
 number = {4},
 pages = {676--685},
 publisher = {Institute of Mathematical Statistics},
 volume = {19},
 year = {2004}
}

@article{ge2003resampling,
  title={Resampling-based multiple testing for microarray data analysis},
  author={Ge, Youngchao and Dudoit, Sandrine and Speed, Terence P},
  journal={Test},
  volume={12},
  number={1},
  pages={1--77},
  year={2003},
  publisher={Springer}
}

@article{anastasiou2022detecting,
  title={Detecting multiple generalized change-points by isolating single ones},
  author={Anastasiou, Andreas and Fryzlewicz, Piotr},
  journal={Metrika},
  volume={85},
  number={2},
  pages={141--174},
  year={2022},
  publisher={Springer}
}

@article{genomics1,
    author = {Cao, Hongyuan and Biao Wu, Wei},
    title = "{Changepoint estimation: another look at multiple testing problems}",
    journal = {Biometrika},
    volume = {102},
    number = {4},
    pages = {974-980},
    year = {2015},
    doi = {10.1093/biomet/asv031},
}

@article{genomics2,
    author = {Jia, Shengji and Shi, Lei},
    title = {Efficient change-points detection for genomic sequences via cumulative segmented regression},
    journal = {Bioinformatics},
    volume = {38},
    number = {2},
    pages = {311-317},
    year = {2021},
    doi = {10.1093/bioinformatics/btab685},
}

@article{seeded_bs,
	title = {Seeded {B}inary {S}egmentation: {A} general methodology for fast and optimal changepoint detection},
	volume = {110},
	shorttitle = {Seeded binary segmentation},
	doi = {10.1093/biomet/asac052},
	number = {1},
	journal = {Biometrika},
	author = {Kovács, S and Bühlmann, P and Li, H and Munk, A},
	year = {2023},
	pages = {249--256},
}

@Article{seismic1,
AUTHOR = {Piana Agostinetti, N. and Sgattoni, G.},
TITLE = {Changepoint detection in seismic double-difference data: application of a trans-dimensional algorithm to data-space exploration},
JOURNAL = {Solid Earth},
VOLUME = {12},
YEAR = {2021},
NUMBER = {12},
PAGES = {2717--2733},
DOI = {10.5194/se-12-2717-2021}
}

@article{seismic2,
    author = {Touati, Sarah and Naylor, Mark and Main, Ian},
    title = {Detection of change points in underlying earthquake rates, with application to global mega-earthquakes},
    journal = {Geophysical Journal International},
    volume = {204},
    number = {2},
    pages = {753-767},
    year = {2015},
    doi = {10.1093/gji/ggv398},
}

@article{astronomy2,
  title={Gamma-ray burst detection with Poisson-FOCuS and other trigger algorithms},
  author={Dilillo, Giuseppe and Ward, Kes and Eckley, Idris A and Fearnhead, Paul and Crupi, Riccardo and Evangelista, Yuri and Vacchi, Andrea and Fiore, Fabrizio},
  journal={The Astrophysical Journal},
  volume={962},
  number={2},
  pages={137},
  year={2024},
  publisher={IOP Publishing}
}

@article{Cho2012,
 author = {Haeran Cho and Piotr Fryzlewicz},
 journal = {Statistica Sinica},
 number = {1},
 pages = {207--229},
 publisher = {Institute of Statistical Science, Academia Sinica},
 title = {Multiscale and multilevel technique for consistent segmentation of nonstationary time series},
 volume = {22},
 year = {2012}
}

@article{DAIS,
  title={Data-adaptive structural change-point detection via isolation},
  author={Anastasiou, Andreas and Loizidou, Sophia},
  journal={Statistics and Computing},
  volume={35},
  number={5},
  pages={117},
  year={2025},
  publisher={Springer}
}

@article{WBS,
  title={Wild binary segmentation for multiple change-point detection},
  author={Fryzlewicz, Piotr},
  journal={The Annals of Statistics},
  volume={42},
  number={6},
  pages={2243--2281},
  year={2014},
  publisher={Institute of Mathematical Statistics},
}

@article{NOT,
  title={{N}arrowest-Over-Threshold Detection of Multiple Change Points and Change-Point-Like Features},
  author={Baranowski, Rafal and Chen, Yining and Fryzlewicz, Piotr},
  journal={Journal of the Royal Statistical Society: Series B (Statistical Methodology)},
  volume={81},
  number={3},
  pages={649--672},
  year={2019},
  doi={https://doi.org/10.1111/rssb.12322},
  publisher={Wiley Online Library},
}

@inproceedings{binary_segmentation,
  title={Detecting “disorder” in multidimensional random processes},
  author={Vostrikova, Lyudmila Yur'evna},
  booktitle={Doklady Akademii Nauk},
  volume={259},
  pages={270--274},
  year={1981},
  organization={Russian Academy of Sciences}
}

@article{WBS2,
  title={Detecting possibly frequent change-points: {Wild Binary Segmentation} 2 and steepest-drop model selection},
  author={Fryzlewicz, Piotr},
  journal={Journal of the Korean Statistical Society},
  volume={49},
  number={4},
  pages={1027--1070},
  year={2020},
  publisher={Springer}
}

@article{review2021,
title = {Selective review of offline change point detection methods},
journal = {Signal Processing},
volume = {167},
pages = {107299},
year = {2020},
doi = {https://doi.org/10.1016/j.sigpro.2019.107299},
author = {Charles Truong and Laurent Oudre and Nicolas Vayatis},
}

@techreport{stephens1969techniques,
  title={Techniques for directional data},
  author={Stephens, Michael A},
  year={1969}
}

@article{wind_direction,
title = {New methods to assess wind resources in terms of wind speed, load, power and direction},
journal = {Renewable Energy},
volume = {129},
pages = {168-182},
year = {2018},
doi = {https://doi.org/10.1016/j.renene.2018.05.088},
author = {G.K. Gugliani and A. Sarkar and C. Ley and S. Mandal},
}

@article{gill2010circular,
  title={Circular data in political science and how to handle it},
  author={Gill, Jeff and Hangartner, Dominik},
  journal={Political Analysis},
  volume={18},
  number={3},
  pages={316--336},
  year={2010},
  publisher={Cambridge University Press}
}

@book{ley_modern_2017,
	address = {Boca Ratón, Florida},
	title = {Modern {Directional} {Statistics}},
	isbn = {978-1-4987-0664-3},
	publisher = {Chapman and Hall/CRC Press},
	author = {Ley, C. and Verdebout, T.},
	year = {2017},
}

@misc{loizidou2025modellingtoroidalcylindricaldata,
      title={Modelling toroidal and cylindrical data via the trivariate wrapped {Cauchy} copula with non-uniform marginals}, 
      author={Sophia Loizidou and Christophe Ley and Shogo Kato and Kanti V. Mardia},
      year={2025},
      eprint={2511.10336},
      archivePrefix={arXiv},
      primaryClass={stat.ME},
    note={arXiv:2511.10336}
}

@article{lagona_hidden_2015,
	title = {A hidden {M}arkov model for the analysis of cylindrical time series},
	volume = {26},
	rights = {Copyright © 2015 John Wiley \& Sons, Ltd.},
	doi = {10.1002/env.2355},
	pages = {534--544},
	number = {8},
	journal = {Environmetrics},
	author = {Lagona, Francesco and Picone, Marco and Maruotti, Antonello},
	urldate = {2023-11-13},
	year = {2015},
	langid = {english},
}

@Manual{circular,
    title = {{R} package \texttt{circular}: Circular Statistics
      (version 0.5-2)},
    author = {Claudio Agostinelli and Ulric Lund},
    year = {2025},
    url = {https://CRAN.R-project.org/package=circular},
  }

@article{hawkins2015segmentation,
  title={Segmentation of circular data},
  author={Hawkins, Douglas M and Lombard, F},
  journal={Journal of Applied Statistics},
  volume={42},
  number={1},
  pages={88--97},
  year={2015},
  publisher={Taylor \& Francis}
}

@article{vsidak1967rectangular,
  title={Rectangular confidence regions for the means of multivariate normal distributions},
  author={{\v{S}}id{\'a}k, Zbyn{\v{e}}k},
  journal={Journal of the American statistical association},
  volume={62},
  number={318},
  pages={626--633},
  year={1967},
  publisher={Taylor \& Francis}
}

@article{seismic3,
    author = {Ma, Yiming and Anastasiou, Andreas and Montiel, Fabien},
    title = {Automated detection of short-term slow slip events using {GNSS} data via change-point analysis},
    journal = {Geophysical Journal International},
    volume = {244},
    number = {3},
    pages = {ggaf517},
    year = {2025},
    doi = {10.1093/gji/ggaf517},
}

@book{LV19,
	address = {Boca Rat{\'o}n, Florida},
	editor = {Ley, C. and Verdebout, T.},
	publisher = {Chapman and Hall/CRC Press},
	title = {Applied Directional Statistics: Modern Methods and Case Studies},
	year = {2018}
}

@article{CHO202476,
title = {Data segmentation algorithms: Univariate mean change and beyond},
journal = {Econometrics and Statistics},
volume = {30},
pages = {76-95},
year = {2024},
doi = {https://doi.org/10.1016/j.ecosta.2021.10.008},
author = {Haeran Cho and Claudia Kirch},
}





% \end{thebibliography}

\end{document}